\documentclass[fleqn,usenatbib]{mnras}

\usepackage[utf8]{inputenc}
\usepackage{threeparttable}
\usepackage{amsmath}
\usepackage{graphicx}
\usepackage{epstopdf}
\usepackage{threeparttable}
\usepackage{graphicx, color}
\usepackage{bm}
\usepackage{soul}
\usepackage{amssymb}
\usepackage{booktabs}

\newcommand{\eqb}{\begin{eqnarray}}
\newcommand{\eqe}{\end{eqnarray}}

\defcitealias{fernandez2018}{F19}
\newcommand{\mdotout}{\ensuremath{\dot M_{\rm out}}}
\newcommand{\mdotaccr}{\ensuremath{\dot{M}_{\rm accr}}}

\newcommand{\harm}{\texttt{HARM}}

\newcommand{\harmpi}{\texttt{HARMPI}}

\usepackage[usenames,dvipsnames,svgnames,table]{xcolor}
\definecolor{darkblue}{rgb}{0.0,0.0,0.3}
\hypersetup{colorlinks,%
            linkcolor=darkblue,urlcolor=darkblue,
            anchorcolor=darkblue,citecolor=darkblue} 
\usepackage{txfonts}
\DeclareSymbolFont{cmletters}{OML}{cmm}{m}{it}
\DeclareMathSymbol{v}{\mathalpha}{cmletters}{"76}

\title[Magnetic Fields and Neutron Star Merger Discs]
{The Role of Magnetic Field Geometry in the Evolution of Neutron Star Merger Accretion Discs}

\author[Christie et al.]{I.M.~Christie$^1$\thanks{E-mail: ichristi231@gmail.com}, A. Lalakos$^{1}$\thanks{E-mail: lalakos@u.northwestern.edu}, A. Tchekhovskoy$^1$, R. Fern{\'a}ndez$^2$, F. Foucart$^3$, \newauthor
  E. Quataert$^4$, \& D. Kasen$^{4,5}$\\
$^1$ Center for Interdisciplinary Exploration \& Research in Astrophysics (CIERA),
Physics \& Astronomy, Northwestern University, Evanston, IL 60202, USA \\ $^2$ Department of Physics, University of Alberta, Edmonton, AB T6G 2E1, Canada \\
$^3$ Department of Physics and Astronomy, University of New Hampshire, Durham, NH 03824, USA \\
$^4$ Departments of Physics \& Astronomy, and Theoretical Astrophysics Center, University of California, Berkeley, CA 94720, USA\\
$^5$  Nuclear Science Division, Lawrence Berkeley National Laboratory, Berkeley, CA 94720, USA}

\begin{document}

\date{Received.../Accepted...}

\pagerange{\pageref{firstpage}--\pageref{lastpage}} \pubyear{2019}

\maketitle

\label{firstpage}

\begin{abstract}
Neutron star mergers are unique laboratories of accretion, ejection, and r-process nucleosynthesis. 
We used 3D general relativistic magnetohydrodynamic simulations to study the role of the post-merger magnetic geometry in the evolution of merger remnant discs around stationary Kerr black holes. Our simulations fully capture mass accretion, ejection, and jet production, owing to their exceptionally long duration exceeding $4$~s. 
Poloidal post-merger magnetic field configurations produce jets with energies $E_\mathrm{jet}\sim(4{-}30)\times10^{50}$~erg, isotropic equivalent energies $E_\mathrm{iso}\sim(4{-}20)\times10^{52}$~erg, opening angles $\theta_\mathrm{jet}\sim6{-}13^\circ$, and durations $t_j\lesssim1$~s. Accompanying the production of jets is the ejection of $f_\mathrm{ej}\sim30{-}40\%$ of the post-merger disc mass, continuing out to times $> 1$~s.
We discover that a more natural, purely toroidal post-merger magnetic field geometry generates large-scale poloidal magnetic flux of alternating polarity and striped jets. The first stripe, of $E_\mathrm{jet}\simeq2\times10^{48}\,\mathrm{erg}$, $E_\mathrm{iso}\sim10^{51}$~erg, $\theta_\mathrm{jet}\sim3.5{-}5^\circ$, and $t_j\sim0.1$~s, is followed by $\gtrsim4$~s of striped jet activity with $f_\mathrm{ej}\simeq27\%$. The dissipation of such stripes could power the short gamma-ray burst (sGRB) prompt emission. Our simulated jet energies and durations span the range of sGRBs. We find that although the blue kilonova component is initially hidden from view by the red component, it expands faster, outruns the red component, and becomes visible to off-axis observers. In comparison to GW~170817/GRB~170817A, our simulations under-predict the mass of the blue relative to red component by a factor of few. Including the dynamical ejecta and neutrino absorption may reduce this tension.
\end{abstract}


\section{Introduction}
\label{sec:intro}
\begin{table*}
\centering
\caption{Models considered and initial parameters. From left to right: Maximum magnetic field strength within the initial torus, simulation duration $t_{\rm max}$ in seconds and in units of $r_g /c$, plasma $\beta$ of the initial torus, and simulation resolution in terms of the number of cells (in the radial, poloidal, and toroidal directions), the effective resolution near the midplane $\pi/\Delta\theta$, and cell aspect ratio near the midplane $\Delta r : r\Delta\theta : r\Delta\phi$. Note that the latter two are determined at a radial distance of $40 \, r_g$. }
\begin{threeparttable}
\begin{tabular}{c|c|c|cc|c|c|c|c}
\hline
Model & Field & Max Field & \multicolumn{2}{c|}{Duration, $t_{\rm max}$} & Initial & Total resolution, & Effective $\theta$-re- & Cell aspect ratio,\\
Name & Geometry & Strength (G) & $(s)$ & $(10^5  \, r_g/c)$ & plasma $\langle\beta\rangle$ & $N_r\times N_\theta \times N_\phi $ & solution, $\pi/\Delta\theta$ &$\Delta r : r\Delta\theta : r\Delta\phi$\\
\hline
BPS & Poloidal & $1.1 \times 10^{14}$ & $9.2$ & $6.2$ & $100$ & $512 \times 256 \times 64$ & $256$ &  $3:1:8$ \\
\hline
BPW & Poloidal & $3.6 \times 10^{13}$ & $4.4$ & $3$ & $850$ & $512 \times 256 \times 128$ & $640$ & $8:1:10$\\
\hline
BT & Toroidal & $4.7 \times 10^{14}$ & $4.3$ & $2.9$ & $5$ & $512 \times 256 \times 128$ & $256$ & $3:1:4$\\
\hline
\end{tabular}
\end{threeparttable}
\label{table:model_resolutions}
\end{table*}
The recent detection of the neutron star (NS) merger GW170817 in both gravitational and electromagnetic (EM) waves has marked a monumental achievement for multi-messenger astronomy \citep{abbott2017b,abbott2017a}. The first detected component of the EM signal presented itself as a short duration $\gamma$-ray burst (GRB), detected $\sim 1.7$~s after the merger, lasting for $\sim 2$~s (GRB 170817A) \citep{goldstein2017,savchenko2017}. This detection provides strong supporting evidence that NS mergers are the progenitors for short GRBs \citep{blinnikov2984,paczynski1986,eichler1989,narayan1992}. The prompt and afterglow emission associated with this event is believed to be produced by a highly relativistic jet launched by a compact object, either a NS or a black hole (BH) \citep{metzger2012,berger2013}. Moreover, computational studies, using 3D general relativistic magnetohydrodynamic (GRMHD) simulations of BH accretion, in the context of NS mergers, have shown that a naturally forming, laterally non-uniform, structured jet can reproduce \citep{kathirgamaraju2018} the observed radio and X-ray afterglow emission from GW 170817/GRB 170817A \citep{Alexander2018,Margutti2018}.

From this event, it is now widely believed that NS mergers are an important site for r-process nucleosynthesis in the Universe \citep{kasen2017,cote2018,hotokezaka2018}. Confirmation of this fact comes from the photometric and spectroscopic observations of the kilonova \citep{cowperthwaite2017,chornock2017,drout2017,tanaka2017,tanvir2017}, argued to be produced from mildly relativistic (i.e. speeds of $v \sim 0.1 \, c$), neutron-rich ejecta radioactively heated by r-process elements \citep{Metzger2010,roberts2011,tanaka2016,metzger2017}. The kilonova transient was observed to transition from a blue optical component to an infrared one in a few days \citep[][see references therin]{chornock2017}, which is consistent with theory if considering both blue emission, from low-opacity light r-process elements, and red emission, from high-opacity heavy r-process elements.

There are two main mechanisms responsible for mass ejection in the kilonova. The first is through dynamical ejecta being expelled on $\sim$~ms timescales by tidal forces \citep{Rosswog1999,Hotokezaka2013} or shock interactions \citep{Oechslin2007,Sekiguchi2016}. 
The second mechanism involves outflows from an accretion disc formed from bound merger material. 
This disc can evolve on longer timescales (i.e. $\sim 100$~ms $- 1$~s) and expand viscously due to the magnetorotational instability (MRI, \citealt{bal91}), which can also power the accretion that transports the large-scale magnetic flux toward the black hole, leading to relativistic jets.  

Our understanding of the temporal evolution of NS merger accretion discs was previously set by axisymmetric, hydrodynamic simulations (e.g. \citealt{fernandez2013,just2015,Fujibayashi2017}). 
\cite{siegal2017,siegel2018} have presented global 3D GRMHD simulations that tracked the evolution of the accretion disc for $\sim 0.4$~s while including the relevant physical process, e.g. alpha-particle recombination and neutrino cooling, and resolving the MRI. \citet[][hereafter \citetalias{fernandez2018}]{fernandez2018} presented the first post-merger remnant disc simulations longer than a second. Their duration of 9 seconds allowed the vast majority of the merger remnant disc to either accrete or fly out as an outflow. 
This work also saw the formation of relativistic jets and mildly relativistic outflows, in the form of disc winds, with speeds $v \gtrsim 0.25c$, above the upper limit found in \citet{siegal2017,siegel2018}.

In NS mergers, a torus with a primarily toroidal field is expected\footnote{Even though the magnetic field is expected to be toroidally dominated, the geometry may be more complex, containing small-scale orientation flips.} from the tidal disruption of one (or both) neutron stars and from flux freezing (e.g. \citealt{Kiuchi2014}). Traditionally, GRMHD simulations of black hole accretion discs have used poloidal flux loops to initialize the disc, which is known to launch relativistic jets and drive outflows \citep{blanford1977}. 
Several studies have found that while a purely toroidal seed magnetic field is sufficient for the MRI to operate in the disc, such systems produce extremely weak jets \citep{beckwith2008,McKinney2012}.
Recently, \citet{liska_dynamo_2018} demonstrated, for a geometrically thick and radially extended accretion disc, that an initially toroidal magnetic field can
generate a large-scale poloidal magnetic flux through what appears to be an
$\alpha{-}\Omega$ dynamo \citep{moffatt1978} and produce a very powerful jet of power comparable to (or even exceeding) the accretion power.
In fact, if the results of \citet{liska_dynamo_2018} applied to less radially extended post-merger accretion discs, they would imply jets that are $4{-}5$ orders of magnitude too powerful to be consistent with short GRB observations \citep[see, e.g.,][]{Fong2015}. This raises an important question: is the more natural, toroidal post-merger magnetic field geometry even capable of leading to jets of power consistent with GRB observations? More generally, how do the properties of the jets and disc outflows in the aftermath of a binary NS merger depend on the initial post-merger magnetic field geometry?

Here, we perform the first quantitative analysis of how the results of 3D GRMHD simulations of NS merger accretion discs depend upon the initial post-merger magnetic field configuration.
We explore what effect this configuration (e.g. purely poloidal and purely toroidal geometries) has on the accretion rate, relativistic jets, and the large-scale outflows, including the implications for and connections with the observed kilonova of GW 170817/GRB 170817A. In Sec.~\ref{sec:setup}, we briefly describe the simulation setup. In Sec.~\ref{sec:accretion_outflows}, we present our results for the mass rate and energetics of all outflows, including the relativistic jet. In Sec.~\ref{sec:discussion}, we discuss the connection of our results with sGRB observations and the observed kilonova of GW 170817/GRB 170817A while concluding in Sec.~\ref{sec:summary}. 

\begin{table*}
\centering
\caption{Summary of our results. From left to right: Cumulative jet energy $E_{\rm jet}$, cumulative isotropic-equivalent jet energy $E_{\rm iso}$, jet opening angle $\langle \theta_{\rm jet} \rangle$ (averaged over both jets and up to $1$~s), accreted mass $M_{\rm accr}$, ejected mass $M_{\rm ejec}$, ejected mass within the red kilonova component $M_{\rm ejec, red}$ (with electron fraction $Y_{\rm e} < 0.25$) and the blue component $M_{\rm ejec, blue}$ ($Y_{\rm e} > 0.25$), the average radial speed of all ejecta $\langle v_{r} \rangle$, the average radial speed within the red $\langle v_{r}\rangle_{\rm red}$ and blue $\langle v_{r}\rangle_{\rm blue}$ kilonova components, and the average electron fraction $\langle Y_{\rm e}\rangle$ of all ejecta. All mass values listed as percentages are normalized to the initial torus mass ($0.033 \, M_{\odot}$) while speeds are normalized to the speed of light. }
\begin{tabular}{@{\;}c@{\;}|@{\;}cc@{\;}|@{\;}c|@{\;}cc@{\;}|@{\;}cc|@{\;}cc@{\;}|@{\;}cc@{\;}|@{\;}c@{\;\;\;}c@{\;}c@{\;}|@{\;}c@{}}
\hline
Model  & $E_{\rm jet}$ & $E_{\rm iso}$ & $\langle \theta_{\rm jet} \rangle$ & \multicolumn{2}{c@{\;}|@{\;}}{$M_{\rm accr}$} & \multicolumn{2}{c@{\;}|@{\;}}{$M_{\rm ejec}$} & \multicolumn{2}{c@{\;}|@{\;}}{$M_{\rm ejec, red}$} & \multicolumn{2}{c@{\;}|@{\;}}{$M_{\rm ejec, blue}$}  & $\langle v_{r} \rangle$ & $\langle v_{r}\rangle_{\rm red}$ & $\langle v_{r}\rangle_{\rm blue}$ & $\langle Y_{\rm e}\rangle$\\
Name & ($10^{50}$~erg) & ($10^{52}$~erg) & ($^\circ$) & ($\%$) & ($10^{-2} \, M_{\odot}$) & ($\%$) & ($10^{-2} \, M_{\odot}$) & ($\%$) & ($10^{-2} \, M_{\odot}$) & ($\%$) & ($10^{-2} \, M_{\odot}$) & &  & & \\
\hline
BPS & $25$ & $22$ & $13$ & $60$ & $2$ & $40$ & $1.3$ & $37$ & $1.2$ & $3$ & $0.1$ & $0.18$ & $0.17$ & $0.3$ & $0.16$ \\
\hline
BPW & $3.9$ & $3.6$ & $6.4$ & $67$ & $2.2$ & $30$ & $0.99$ & $27$ & $0.89$ & $3$ & $0.1$ & $0.08$ & $0.07$ & $0.16$ & $0.19$ \\
\hline
BT & $0.2$ & $1.3$ & $4.6$ & $71$ & $2.3$ & $27$ & $0.89$ & $25$ & $0.83$ & $2$ & $0.066$ & $0.05$ & $0.05$ & $0.08$ & $0.18$ \\
\hline
\end{tabular}
\label{table:model_results}
\end{table*}

\section{Simulation Setup}
\label{sec:setup}

We performed simulations using \harmpi{}\footnote{\url{https://github.com/atchekho/harmpi}}, an enhanced version of the serial open-source code \harm{} \citep{gammie2003,noble2006}, with the addition of several physical processes, e.g. neutrino cooling and nuclear recombination (for more details, see \citetalias{fernandez2018}). Throughout, we use spherical polar coordinates $r$, $\theta$, $\phi$ in the Kerr-Schild foliation. 
For neutrino cooling, we adopt the emission rates described in \citet{Janka2001} and suppress emission in optically thick regions by a factor of $e^{-\tau_\nu}$, where 
\eqb
\label{eqn:tau_nu}
\tau_\nu = \rho/10^{11} g \, {\rm cm}^{-3},
\eqe
and $\rho$ is the gas density. 
Simulations were initialized with a BH of mass $M_{\rm BH} = 3 \, M_\odot$, where $M_\odot$ is the solar mass, and spin parameter $a = 0.8$, surrounded by a torus of mass $0.033 \, M_\odot$\footnote{We note that these simulations, with a torus mass of $0.033 \, M_\odot$, were performed before the results of GW 170817/GRB 170817 A were announced and observational modeling was performed. Studies have since inferred an initial torus mass of $\sim 0.1 \, M_\odot$ \citep{Shibata2017}. } and constant initial electron fraction $Y_{\rm e} = 0.1$. We employ an ideal gas law equation of state (EOS) with a constant adiabatic index $\gamma_{\rm ad} = 4/3$, where the gas temperature $T$ is determined from the total pressure, with contributions from the radiation, electron, proton, and neutron components:
\eqb
\label{eqn:total_pressure}
P = [1 + Y_{\rm e}] \frac{\rho \, k \, T}{m_{\rm n}} + \frac{1}{3} a_{\rm rad} T^4.
\eqe
Here, $a_{\rm rad}$ is the radiation constant and $m_{\rm n}$ is the neutron mass. The electron fraction $Y_e$ is evolved according to the numerical procedures outlined in \citetalias{fernandez2018}. We note that our choice for the adiabatic index $\gamma_{\rm ad}$ was selected by comparing with hydrodynamic simulations which use a physical EOS (see Appendix~A1 of \citetalias{fernandez2018}).

We performed three simulations differing  only in the initial post-merger magnetic field geometry within the torus. We considered two models, one with a strong poloidal magnetic field configuration (BPS, described in detail in \citetalias{fernandez2018}) and one with a weak field configuration (BPW model). The initial conditions for both models 
are described by a vector potential $A_{\phi} \propto r^{5} \rho^{2}$, which is then modified to maximize the magnetic flux in the torus as described in \citet{Tchekhovskoy2011}. 
For each of the two poloidal configurations, we normalized the magnetic field strength such that the density-weighted ratio of gas to magnetic pressure within the disc,
\eqb
\label{eqn:plasma_beta}
\langle \beta \rangle_\rho = \frac{\int \rho \, p_{\rm gas} \, {\rm d}V}{\int \rho \, p_{\rm mag} \, {\rm d}V},
\eqe
is $\langle \beta \rangle_\rho = 100$ for BPS and $850$ for BPW, respectively. Here ${\rm d}V = \sqrt{-g}\,{\rm d}r \, {\rm d}\theta \, {\rm d}\phi$ is the volume element and $g$ is the determinant of the metric. 
For BPS, the MRI is easily resolved at a moderate resolution throughout the torus and yet the magnetic field is not too strong to violently distort the torus after being amplified by the shear and the MRI. For BPW, the magnetic field is $\sim 3$ times weaker, which requires us to use a numerical grid which is more finely concentrated near the equatorial plane to resolve the MRI well and to use twice as a high resolution in the $\phi$-direction as in BPS. We provide a summary of each configuration setup, including the adopted simulation resolution, in Table~\ref{table:model_resolutions}.

The third and final configuration is a toroidal magnetic field model, denoted as model BT, with plasma $\beta \equiv p_{\rm gas}/p_{\rm mag}= 5$ throughout the torus. 
We adopted such a low $\beta$ value because: i) it was feasible to resolve the MRI given the available computational resources and ii) the magnetic pressure is low enough so it does not disrupt the disc. In all simulations, our numerical grid extends from just inside the event horizon to $ \sim 10^5 \, r_g$ in the radial direction and from $0$ to $\pi$ in the $\theta$ and $\phi$-directions.  

We carried the simulations out to $t_{\rm max}\sim (3{-}6) \times 10^5 \, r_g/c \simeq 4{-}9$~s, where $r_g = GM_{\rm BH} / c^{2}$ is the gravitational radius of the BH and $c$ is the speed of light. Along with the BPS model described in \citetalias{fernandez2018}, these are the longest run simulations to date, as measured in the units of $r_g/c$ (e.g. longer than the $2\times 10^5 r_g/c$ duration in \citealt{narayan2012}). This unusually long duration is necessary for mass ejection to complete: the cumulative ejected mass dependence on time flattens out at late times (see Fig.~\ref{fig:m_dot_out}(b)). It is also necessary to capture the jet activity that lasts several seconds (see Fig.~\ref{fig:E_jet}). We provide a summary of our results in Table~\ref{table:model_results} and include videos of each simulation in Supplementary Information.\footnote{\url{https://goo.gl/ct7Htx}: There are two sets of videos. The first contains two panels, with the left and right panels showing the logarithm of density (in g cm$^{-3}$) and the electron fraction $Y_{\rm e}$, respectively, in a vertical slice (see also Fig.~\ref{fig:B_phi_temporal_snapshots}). The second set displays the mass-weighted red (i.e. $Y_{\rm e} < 0.25$ material) and blue (i.e. $Y_{\rm e} < 0.25$ material) kilonova components and the jet (green) at a distance of $r_{\rm out} = 10^9 \, {\rm cm} \approx 2000 \, r_g$ (see also Fig.~\ref{fig:kilonova_snapshots}).}

\section{Simulation Results}
\label{sec:accretion_outflows}

\subsection{Mass Accretion}
\label{sec:mass-accretion}

\begin{figure}
\centering
\includegraphics[height=0.55\textwidth]{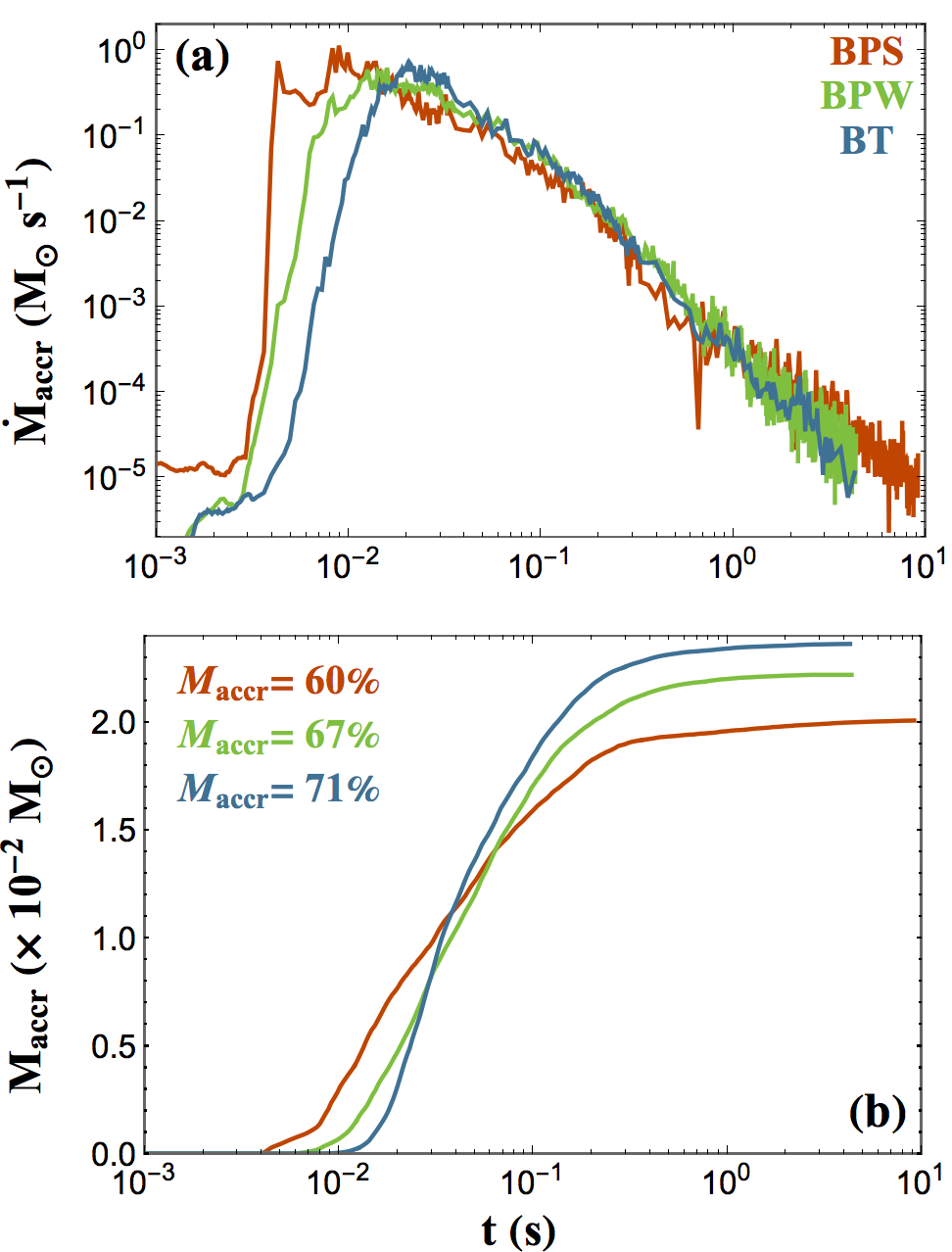}
\caption{For a wide range of post-merger magnetic field geometries (see legend), we find a very similar temporal trend in the rest mass accretion rate (panel a) on the BH such that at late times ($t \gtrsim 5\times 10^{-2}$~s), there is negligible difference between each model. However, these slight differences in the temporal decline of $\dot{M}_{\rm accr}$ imprint themselves as a small variation in the amount of material $M_{\rm accr}$ (panel b) accreted on the BH, with a purely toroidal configuration accreting the most material. The time at which the MRI fully develops coincides with the peak in $\dot{M}_{\rm accr}$. A coloured version of this plot is available online.}
\label{fig:m_dot_combined}
\end{figure}

Upon the start of the simulation, the disc shear leads to the development of the MRI, which amplifies the magnetic field and powers magnetized turbulence in the disc. This drives accretion of gas onto the black hole. As shown in Fig.~\ref{fig:m_dot_combined}(a), the mass accretion rate on the black hole increases and peaks around $10$~ms ($\sim 1000 \, r_g/c$). 
The mass accretion rate peaks slightly earlier for the strong poloidal case and slightly later for weaker magnetic fields. Following the peak, \mdotaccr{} decays in the form of a power-law whose slope is essentially independent of the post-merger field geometry. Interestingly, the power-law decay portion of \mdotaccr{} is roughly the same for all configurations, suggesting that the effects of the magnetic field geometry are not important qualitatively for the evolution of the accretion disc past the initial burn-in period (see also \citealt{beckwith2008}). This decline in the accretion rate comes from the reduction in the mass of the disc, due to both accretion onto the BH and ejection of gas in outflows.

We can perform a more quantitative comparison by looking at the total amount of material accreted by the BH, $M_{\rm accr}$, as shown in Fig.~\ref{fig:m_dot_combined}(b) and Table~\ref{table:model_results}. The amount of accreted material reaches an asymptotic value by $\sim 2$~s for all post-merger geometries. In the strongest poloidal field model, BPS, the BH consumes the least amount of gas, $M_{\rm accr} \sim 60 \%$ ($0.02 \, M_\odot$), followed by $\sim 67 \%$ ($0.022 \, M_\odot$) for weak poloidal field model BPW, and $\sim 71 \%$ ($0.023 \, M_\odot$) for toroidal field model BT. Stronger poloidal magnetic fields lead to stronger outflows, so there is less gas left to be consumed by the BH. Interestingly, the weaker poloidal magnetic field models accrete approximately the same amount of mass but do not reach the hydrodynamic limit (see \citetalias{fernandez2018}). 

\subsection{Relativistic Outflows}
\label{sec:energy-outflows}

\begin{figure}
\centering
\includegraphics[height=0.55\textwidth]{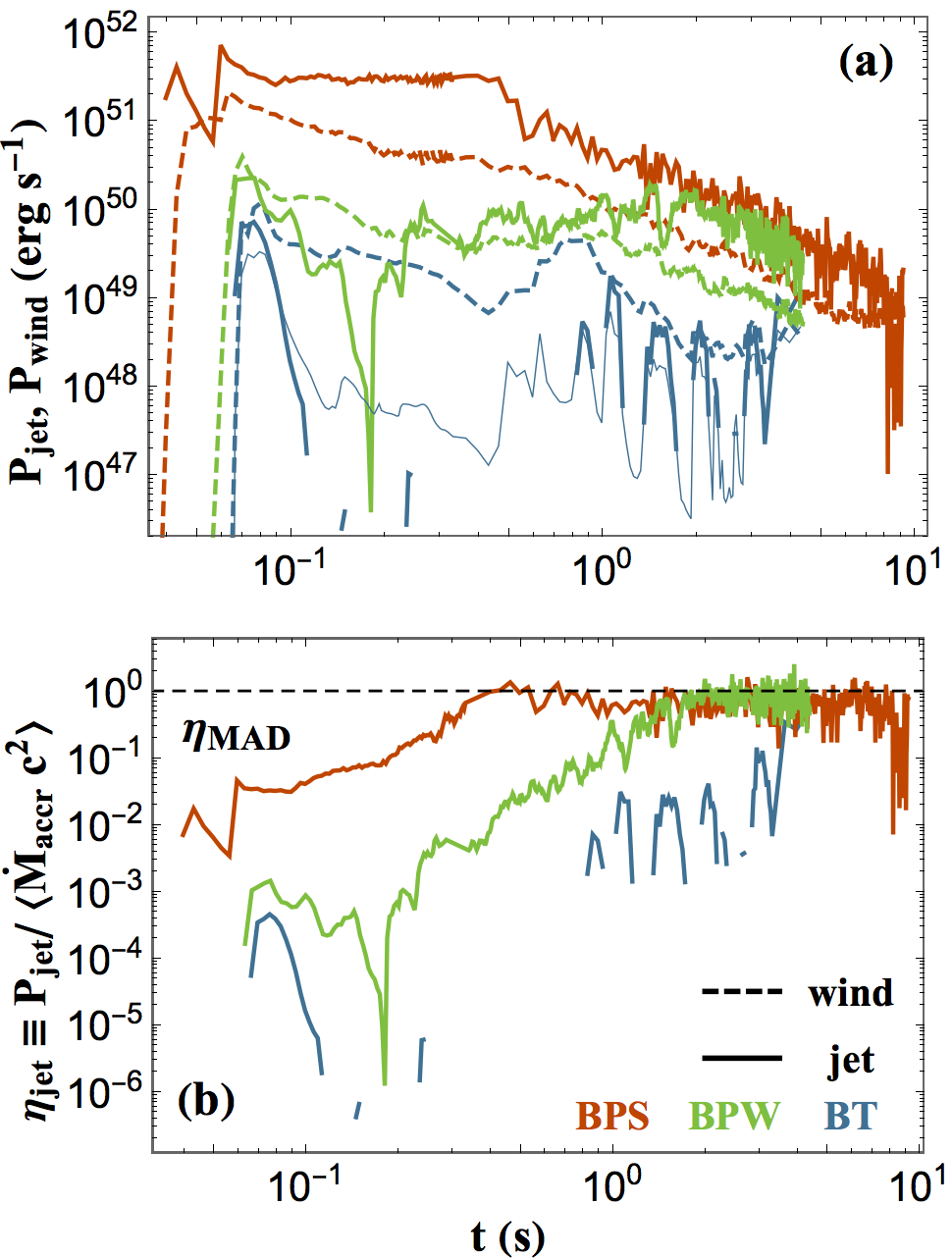}
\caption{The jet power $P_{\rm jet}$ (panel a, solid lines) and jet efficiency $\eta_{\rm jet}$ (panel b) at $r_{\rm out} = 10^9 \, {\rm cm} \approx 2000 \, r_g$ strongly depend upon the post-merger magnetic field geometry (see legend in panel b), producing a $\sim 10^2$ times difference in $P_{\rm jet}$ between the different geometries. For all geometries, $\eta_{\rm jet}$ approaches $\eta_{\rm MAD} \sim 1$ coincidentally with the time the disc reaches its MAD state (see Fig.~\ref{fig:phi_BH}(b)). This is the first demonstration of a powerful jet forming from a purely toroidal magnetic post-merger geometry. In the BT model, the surrounding disc winds disrupt the jet, fueling its intermittence, portrayed as the broken lines in $P_{\rm jet}$ and $\eta_{\rm jet}$. During jet disruptions, the EM power at $r_{\rm out}$ (thin solid line) is therefore solely contained within the surrounding disc winds. The power contained within the disc winds is $\sim$~few times less powerful than the jet and follows a similar temporal trend as $P_{\rm jet}$. A coloured version of this plot is available online.}
\label{fig:e_dot_em}
\end{figure}

The simulated discs can eject energy in the form of outflows launched by the magnetic fields twisted by
the rotation of the BH \citep{blanford1977,2001MNRAS.326L..41K,2010ApJ...711...50T} or the accretion disc \citep{1982MNRAS.199..883B}. Typically, numerical simulations of BH accretion show a combination of the two: BH-powered relativistic jets surrounded by sub-relativistic disc-powered winds \citep{2005ApJ...630L...5M, hawley2006, Tchekhovskoy2011, 2015ASSL..414...45T}. We compute the net sum of these outflow powers through a surface integral
\begin{equation}
\label{eqn:tot_power}
P_{\rm tot} (r) = - \oiint T^r_t\, {\rm d}A,
\end{equation}
where ${\rm d}A = \sqrt{-g} \, {\rm d}\theta \, {\rm d}\phi$ is the area element,
\begin{equation}
  T^\mu_\nu = (\rho+u+P+b^2)u^\mu u_\nu + (P+\frac{1}{2}b^2)\delta^\mu_\nu - b^\mu b_\nu
  \label{eq:stress-energy}
\end{equation}
is the stress-energy tensor, $b^\mu$ is the magnetic four-vector, $b^2 = b^\mu b_\mu$ is twice the magnetic pressure, and $u^\mu$ is the proper velocity  \citep{gammie2003}. To distinguish jets and winds, we make use of the specific energy flux, $\mu = -T^r_t/(\rho u^r)$:
the value of $\mu$ determines the maximum possible Lorentz factor an outflow would achieve if all of its internal and magnetic energy were converted into kinetic energy. We refer to regions with $\mu \ge 2$ as the relativistic jets and $\mu<2$ as the mildly relativistic winds \citep{Tchekhovskoy2011,2015ASSL..414...45T}.

Figure~\ref{fig:e_dot_em}(a) shows the jet and wind powers, $P_{\rm jet}$ and $P_{\rm wind}$ respectively, evaluated at $r_{\rm out}$.
Because relativistic jets are powered by large-scale poloidal magnetic fields, it is perhaps not surprising that the strong poloidal flux model, BPS, forms powerful relativistic jets. In fact, Fig.~\ref{fig:e_dot_em}(a) shows that the jet power ramps up shortly after the light crossing time, $r_{\rm out}/c \simeq 0.033$~s, and flattens out at $P_{\rm jet} \simeq 3\times10^{51}\,\mathrm{erg\,s^{-1}}$ until $\sim0.5$~s (\citetalias{fernandez2018}). How can the jet power remain constant even though the mass accretion rate rapidly declines, as seen in Fig.~\ref{fig:m_dot_combined}? Fig.~\ref{fig:e_dot_em}(b) shows that this decline leads to the increase in jet efficiency, $\eta_{\rm jet} \equiv P_{\rm jet}/ \langle\dot{M}_{\rm accr} \, c^{2}\rangle$ -- the ratio of jet to accretion power -- from $1\%$ at $t\sim0.05$~s to $100\%$ at $0.5$~s in our BPS model.

\begin{figure}
\centering
\includegraphics[height=0.55\textwidth]{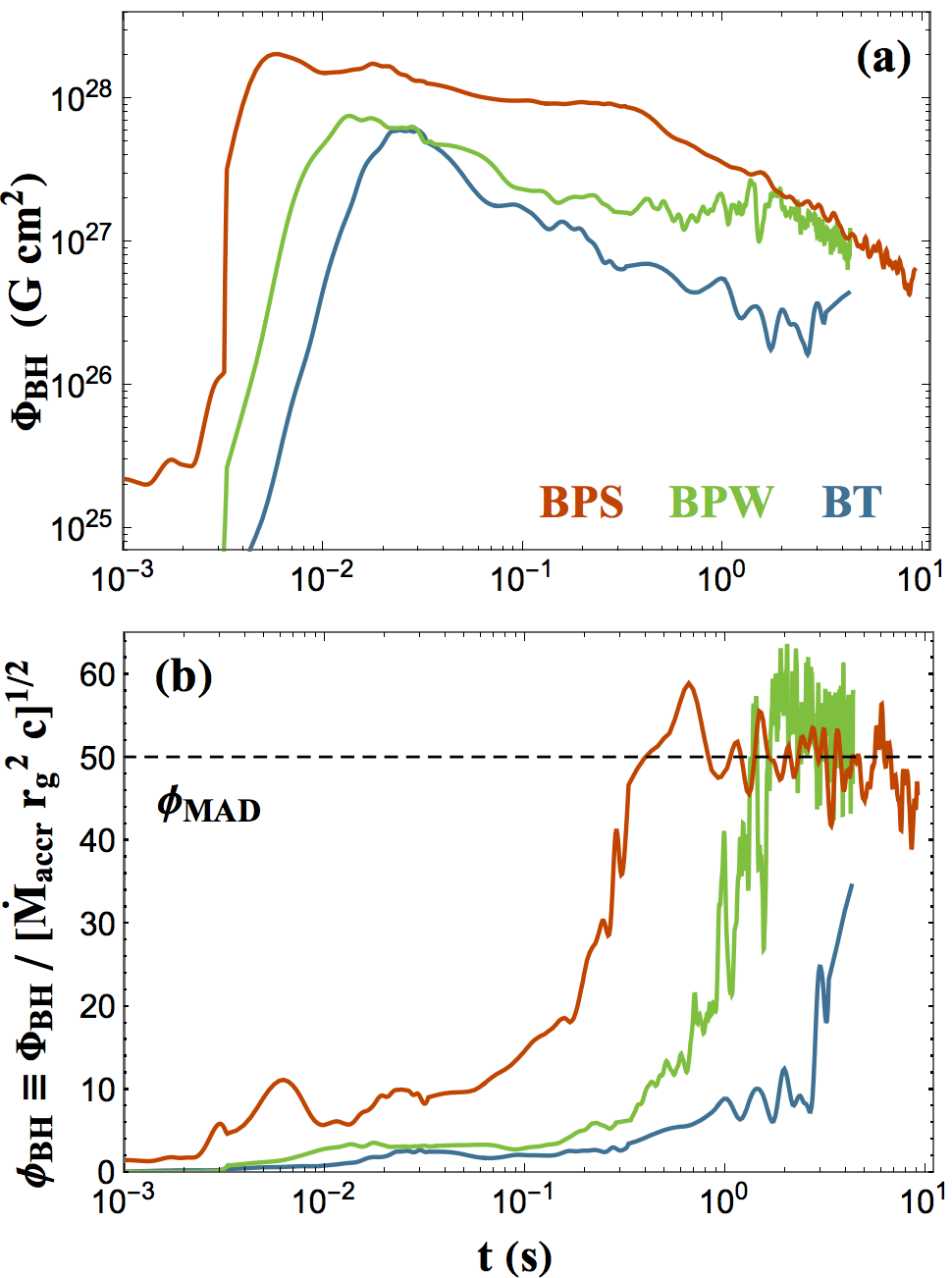}
\caption{The physical $\Phi_{\rm BH}$ (panel a) and normalized $\phi_{\rm BH} \equiv \Phi_{\rm BH}/ (\dot{M}_{\rm accr} r_g^2 c)^{1/2}$ (panel b) poloidal magnetic flux on the BH, powering the relativistic jets, varies significantly with the post-merger field geometry (see legend in panel a). For all configurations, $\phi_{\rm BH}$ eventually reaches or approaches a critical value of $\phi_{\rm MAD} \sim 50$, demonstrating that in the context of short GRBs, large-scale poloidal magnetic flux on the BH can become dynamically-important and lead to the development of a MAD (see eqn.~\ref{eq:Phibh_mad}). A coloured version of this plot is available online.}
\label{fig:phi_BH}
\end{figure}
 
This change in efficiency by two orders of magnitude implies that, unlike a typical expectation that jet power follows mass accretion rate, there is \emph{no one-to-one connection between the mass accretion rate and jet power}. 
To understand this, it is helpful to look at the behavior of the large-scale poloidal magnetic flux that passes through the BH and powers the relativistic jets,
\begin{equation}
\label{eqn:B_phi_BH}
\Phi_{\rm BH} = 0.5 \int_{r = r_{\rm H}} |B^{r}| \, d{\rm A},
\end{equation}
where the integral is over both hemispheres of the event horizon, $r_{\rm H} = r_g [1 + \sqrt{1- a^2}]$, 
and the factor of $0.5$ converts it to a single hemisphere \citep{Tchekhovskoy2011}. Because the jet power is proportional to the square of BH magnetic flux \citep{blanford1977,2010ApJ...711...50T},
\begin{equation}
  P_{\rm jet} \propto \Phi_{\rm BH}^2,
  \label{eq:pjet-phibhsq}
\end{equation}
the constancy of the jet power would imply the constancy of $\Phi_{\rm BH}$ \citep{Tchekhovskoy2015}. Fig.~\ref{fig:phi_BH}(a) shows that within $\sim 10$~ms after the start of the simulation, the central BH receives most of the large-scale magnetic flux available after the merger, after which the BH flux indeed remains approximately constant (to within a factor of $2$) until $t\sim 0.5$~s. This near-constancy of BH magnetic flux results in a near-constancy of jet power.

\begin{figure}
\centering
\includegraphics[height=0.285\textwidth]{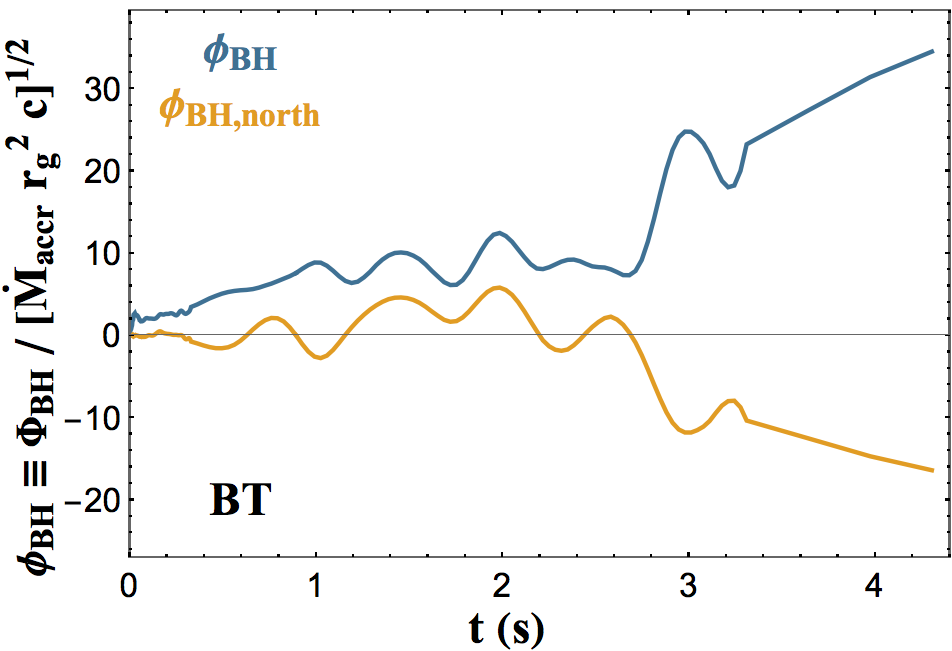}
\caption{The time-dependence of normalized poloidal magnetic flux on the northern hemisphere of the BH ($\phi_{\rm BH, north}$, yellow) shows flips in the magnetic polarity of jets (see also Fig.~\ref{fig:B_phi_temporal_snapshots}). Namely, the polarity on the northern hemisphere of the BH flips sign several times throughout the simulation, resulting in the production of current sheets that propagate along the jets. 
Note that the total absolute magnetic flux exceeds that through the northern hemisphere, implying the presence of non-equatorial current sheet(s) on the BH event horizon. A colour version of this plot is available online.}
\label{fig:phi_BH_BT}
\end{figure}

However, this magnetic flux cannot remain on the BH indefinitely: by the no-hair theorem \citep{1973grav.book.....M}, the magnetic flux would not be able to stay on the BH after all of the accreting gas is gone. In fact, the mass accretion rate sets an upper limit to the BH magnetic flux: if outward magnetic pressure force exceeds the inward pull of BH gravity on the disc, then the magnetic flux leaves the BH by cutting its way through the surrounding disc \citep{Tchekhovskoy2011,2015ASSL..414...45T,Tchekhovskoy2015}. At this point,
the flow turns into a magnetically arrested disc (MAD; \citealt{narayan2003,Igumenshchev2003}): the magnetic flux on the BH becomes dynamically important and obstructs the accreting gas (see Fig.~\ref{fig:phi_BH}(b)).
In the MAD state, the magnetic flux remains (in a time-average sense) at its maximum value set by the weight of the disc: the magnetic pressure, which is proportional to the square of the BH magnetic flux, $P_{\rm mag}\propto \Phi_{\rm BH}^2$, scales linearly with the disc mass that in turn is proportional to mass accretion rate, $\dot{ M}_{\rm accr}$. As a result, in a MAD, the BH magnetic flux is proportional to the square root of the mass accretion rate,
\begin{equation}
\Phi_{\rm BH}\simeq 50 \, (\dot{M}_{\rm accr} r_g^2 c)^{1/2},\label{eq:Phibh_mad}
\end{equation}
where we have included the proportionality factor \citep{2015ASSL..414...45T}. Equivalently, in the MAD state, the dimensionless BH magnetic flux,
\begin{equation}
    \label{eq:phi_BH_normalized}
    \phi_{\rm BH} = \frac{\Phi_{\rm BH}}{(\dot{M}_{\rm accr} r_g^2 c)^{1/2}},
\end{equation}
is $\simeq50$.
Equations~\eqref{eq:pjet-phibhsq} and \eqref{eq:Phibh_mad} imply that in the MAD state,
\begin{equation}
  P_{\rm jet} \simeq 1.3 a^2 \dot{M}_{\rm accr} c^2,
  \label{eq:Pjet_mad}
\end{equation}
where we again have provided the proportionality factor \citep{2015ASSL..414...45T}. That this numerical factor for our rapidly spinning BH with $a = 0.8$ is of order unity, implies an order unity jet efficiency $\eta_{\rm jet}$, as seen in Fig.~\ref{fig:e_dot_em}(b).  Thus, at late times, the jet power approximately equals the accretion power, and both decay as a power-law in time with a slope of $\sim -2.1$ inferred from fitting $\dot{M}_{\rm accr}$ shown in Fig.~\ref{fig:m_dot_combined}.
\begin{figure*}
\centering
\includegraphics[height=0.34\textwidth]{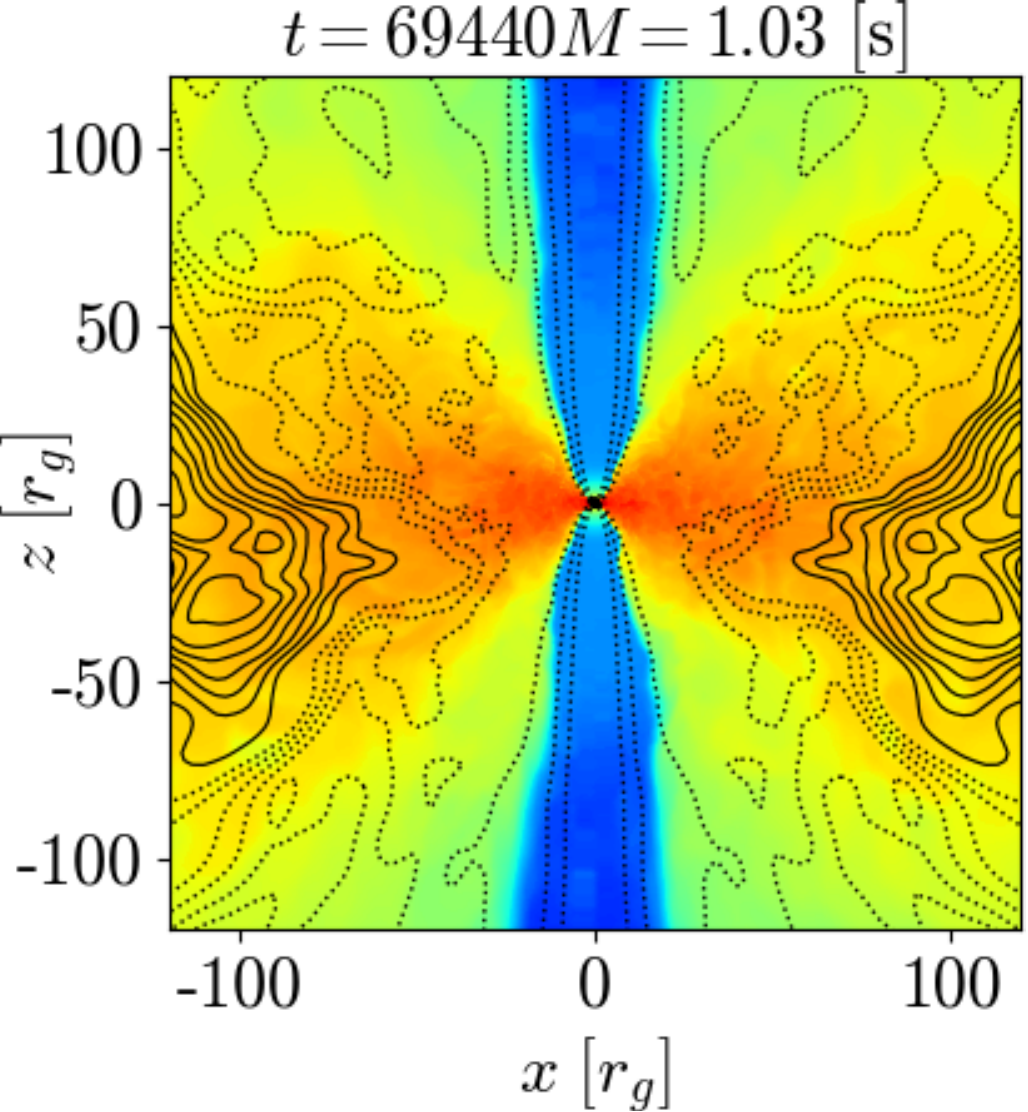}
\hspace{0.125cm}
\includegraphics[height=0.34\textwidth,trim=1.3cm 0cm 0cm 0, clip]{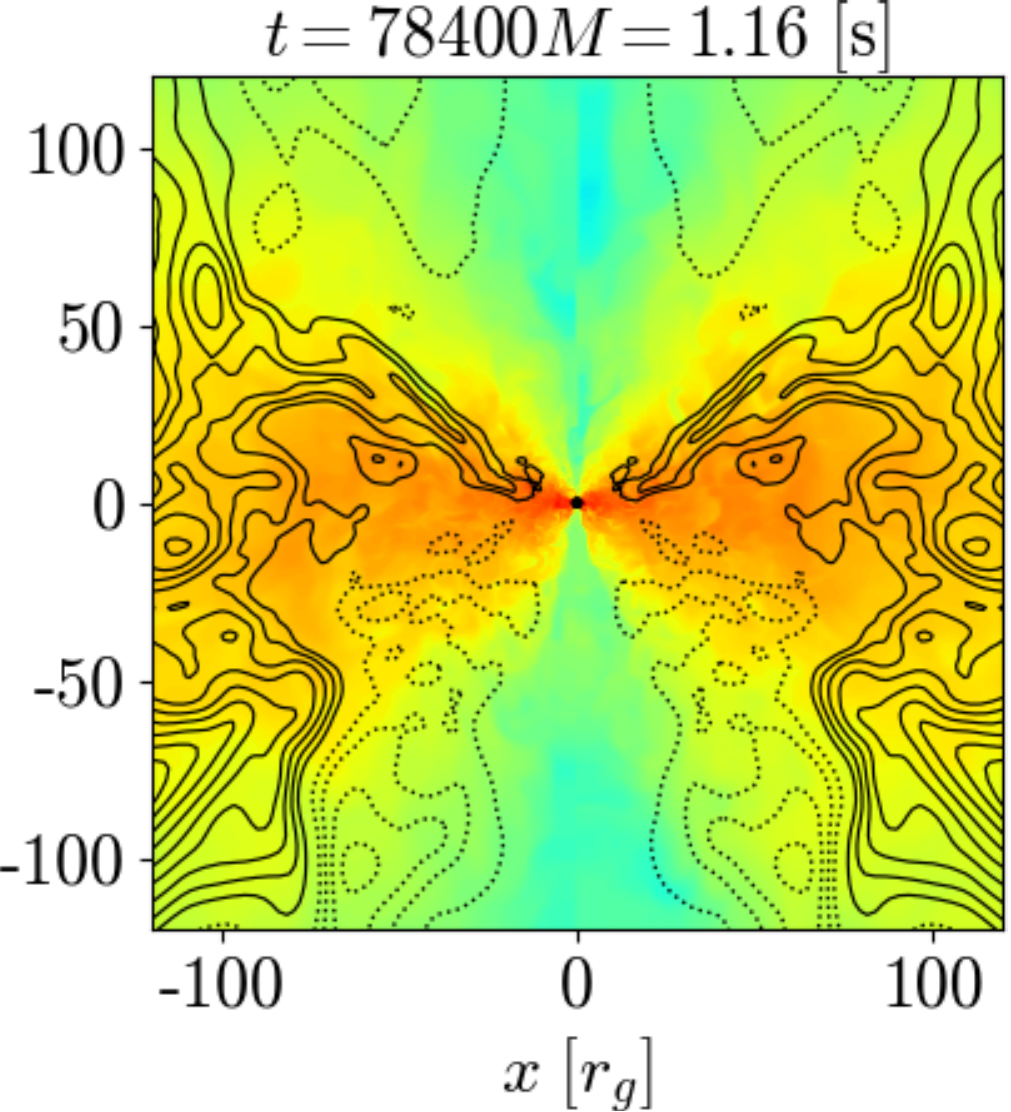}
\hspace{0.125cm}
\includegraphics[height=0.3407\textwidth,trim=1.3cm 0cm 0cm 0, clip]{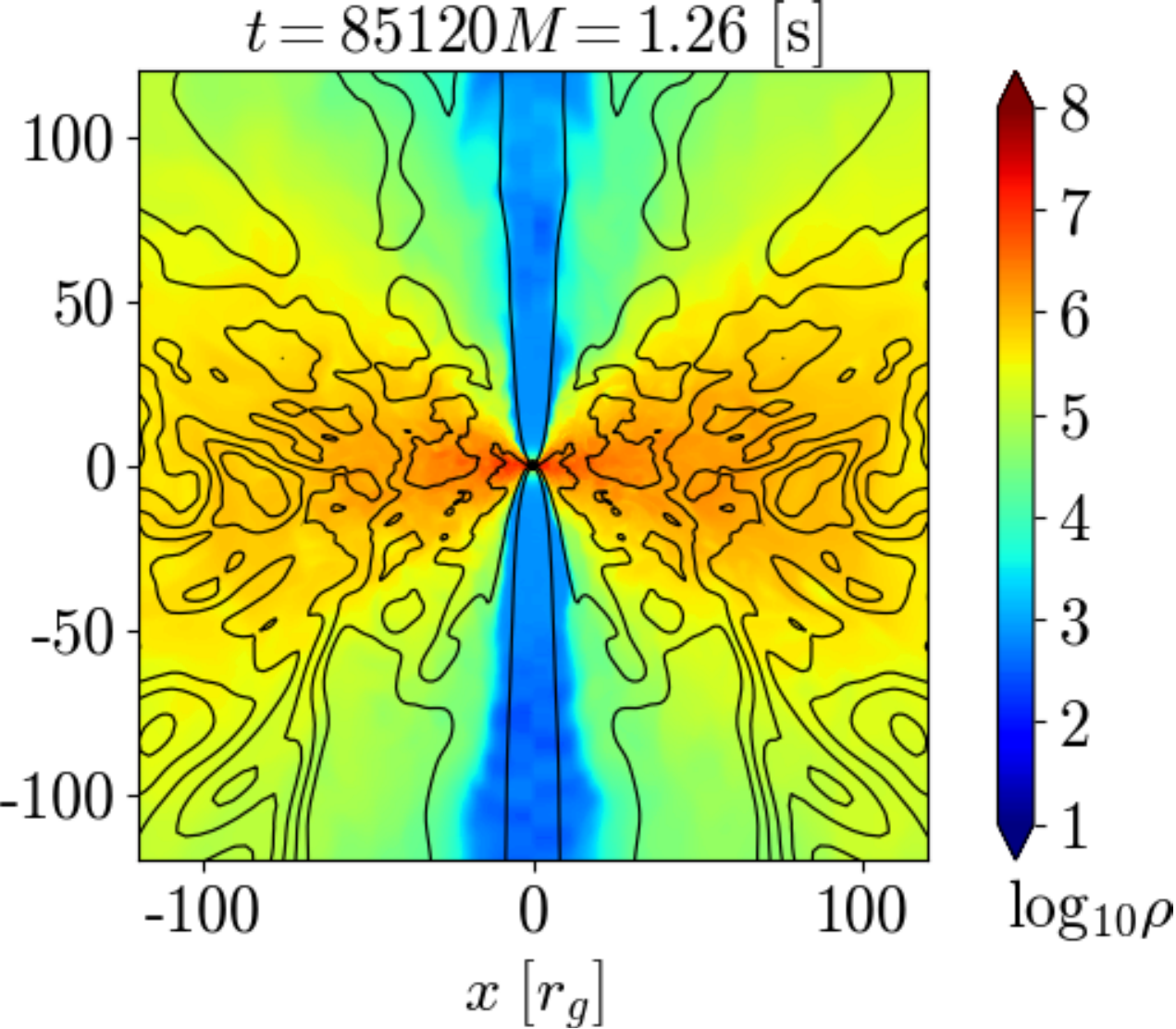}
\caption{A sequence of vertical slices through density in our toroidal post-merger field simulation BT showing flips in magnetic polarity (solid and dotted lines representing positive and negative polarity, respectively) of the jets (blue regions).  The magnetic polarity switches from negative (left panel) to positive (right panel), causing the BH magnetic flux to reconnect away, dropping the power in the relativistic jets (see Fig.~\ref{fig:e_dot_em}), and the surrounding disc winds to choke the jets (middle panel; see also Fig.~\ref{fig:e_dot_em}(a)). This is the first demonstration of a striped jet formation in a BH accretion simulation with the stripes naturally emerging through a large-scale poloidal flux dynamo in the disc rather than introduced through an initial condition. A colour version of this plot is available online.}
\label{fig:B_phi_temporal_snapshots}
\end{figure*}

Our weaker post-merger poloidal field BPW model leads to an order of magnitude weaker jet than in the BPS model: $P_{\rm jet}$ initially peaks at $3\times 10^{50}\,\mathrm{erg\,s^{-1}}$ and eventually plateaus at $\simeq 10^{50}\,\mathrm{erg\,s^{-1}}$. As in BPS, this flattening is due to the constancy of the large-scale poloidal magnetic flux on the BH, as seen in Fig.~\ref{fig:phi_BH}(a) between $t\simeq0.1$ and $2$~s. Because the BH magnetic flux is weaker in this model than in the BPS model, it takes longer for the mass accretion rate to drop to the critical value, $\dot M_{\rm crit}\simeq \Phi_{\rm BH}^2/(50^2 r_g^2 c)$, at which, the magnetic flux becomes dynamically-important, leading to the formation of a MAD.
Beyond this point, the jet power follows the mass accretion rate, $P_{\rm jet} \sim \dot{M}_{\rm accr} \, c^2$, as is typical of MADs and seen in Fig.~\ref{fig:e_dot_em}(b).
Models BPS and BPW demonstrate that compact accretion discs (as typical for short GRBs) can naturally reach a MAD state (see \citealt{Proga2006} for 2D analog). Because the jet power tracks the mass accretion rate, we are conveniently provided with an inside view of the late time accretion on the BH.

Surprisingly, we find that the purely toroidal post-merger geometry BT model also launches jets of substantial power, $P_{\rm jet} \simeq 10^{50}\,\mathrm{erg\,s^{-1}}$ at $t\simeq0.1$~s. At later times, $t\sim 0.5{-}4$~s, the jets ``flicker'' by intermittently switching on and off, with typical $P_{\rm jet}\sim 10^{49}\,\mathrm{erg\,s^{-1}}$. Fig.~\ref{fig:e_dot_em}(b) shows that the corresponding jet efficiency gradually increases in time, eventually approaching $100$\%. We discuss this in more detail in Sec.~\ref{sec:poloidal_flux}.

\subsection{disc Dynamo and Poloidal Magnetic Flux Generation}
\label{sec:poloidal_flux}

How is it possible that even in the absence of any poloidal magnetic flux, model BT produces relativistic jets, which require large-scale poloidal magnetic flux \citep{beckwith2008,McKinney_Bladnford2009,McKinney2012}? This has never been seen in numerical simulations of compact discs, such as those expected in a binary merger. The yellow line in Fig.~\ref{fig:phi_BH_BT} shows the time-dependence of the magnetic flux through the northern hemisphere of the BH, $\phi_{\rm BH,north}$, normalized by the mass accretion rate (see eqn.~\ref{eq:phi_BH_normalized}). While initially starting at zero, the magnetic flux increases in magnitude. This suggests that that the initial purely toroidal post-merger magnetic field undergoes a dynamo-like process that can generate large-scale poloidal magnetic flux. We can see the dynamo action in the movies\footnote{\url{https://goo.gl/ct7Htx}} of model BT through the emergence of poloidal magnetic loops above and below the equatorial plane. This behavior is consistent with an $\alpha{-}\Omega$ like poloidal flux dynamo \citep{moffatt1978,liska_dynamo_2018}.
The dimensionless flux evolves slowly with time, approaching the critical MAD state (although $\phi_{\rm BH} \sim 50$ is not strictly reached in the duration of the simulation). The polarity of the emerging poloidal magnetic flux appears to switch at random, likely reflecting the randomness of the magnetized turbulence underlying the dynamo.

To understand the connection between the magnetic flux and the jet power, it is helpful to look at the absolute magnetic flux, $\phi_{\rm BH}$ (see eqn.~\ref{eq:phi_BH_normalized}), the square of which controls the jet power.
We see that the jet power varies significantly (see Fig.~\ref{fig:e_dot_em}(a)), with the jets shutting off at multiple times. The jets become suppressed due to the winds of the surrounding disc choking and disrupting the jets, especially at the times when the magnetic flux vanishes and the jets are weakened. Fig.~\ref{fig:B_phi_temporal_snapshots} illustrates how one such flux flip happens. The jets seen in blue in the left panel (at $t \approx 1.03$~s), have a well-defined structure. In contrast, in the middle panel (at $t\approx 1.16$~s), the surrounding winds disrupt the jets. The right panel shows that eventually the jets manage to push through (at $t \approx 1.25$~s). Such magnetic flux polarity flips occur frequently and do not appear to show any obvious periodicity, as seen in Fig.~\ref{fig:phi_BH_BT}.  The average duration between the flips appears to increase with the increasing simulation time. We do not see such sign flips in the BPS and BPW models, suggesting that at least on large scales, the initial post-merger magnetic flux  dominates in these models the flux, if any, produced by the dynamo. Note that jet power shut-offs occur more frequently than the polarity flips, indicating that many factors (not just the strength of the magnetic flux but also, e.g., mass-loading of the polar regions by the ambient gas) determine the success of relativistic jet formation.  When the jets are shut off, the EM power at $r_{\rm out}$ is solely contained within the surrounding disc winds. This power, displayed as the thin blue line in Fig.~\ref{fig:e_dot_em}(a), can contribute substantially to the total EM power of the combined jet~$+$~wind regions. 
At late times, $t\gtrsim4$~s, Figs.~\ref{fig:phi_BH}(b) and~\ref{fig:phi_BH_BT} show that $\phi_{\rm BH}$ approaches the critical value of $50$, and Fig.~\ref{fig:e_dot_em}(b) shows that the dimensionless jet power $\eta_{\rm jet}$ approaches the critical value of $1$. This suggests that even absent poloidal post-merger magnetic flux, the system manages to generate its own poloidal flux and approach (but not quite reach) the MAD state. With a longer simulation, it is plausible that we would see a full MAD state develop in our BT model, allowing us to use the jet power as an observational window in the accretion on the BH (see eqn.~\ref{eq:Pjet_mad}).
\begin{figure}
\centering
\includegraphics[height=0.55\textwidth]{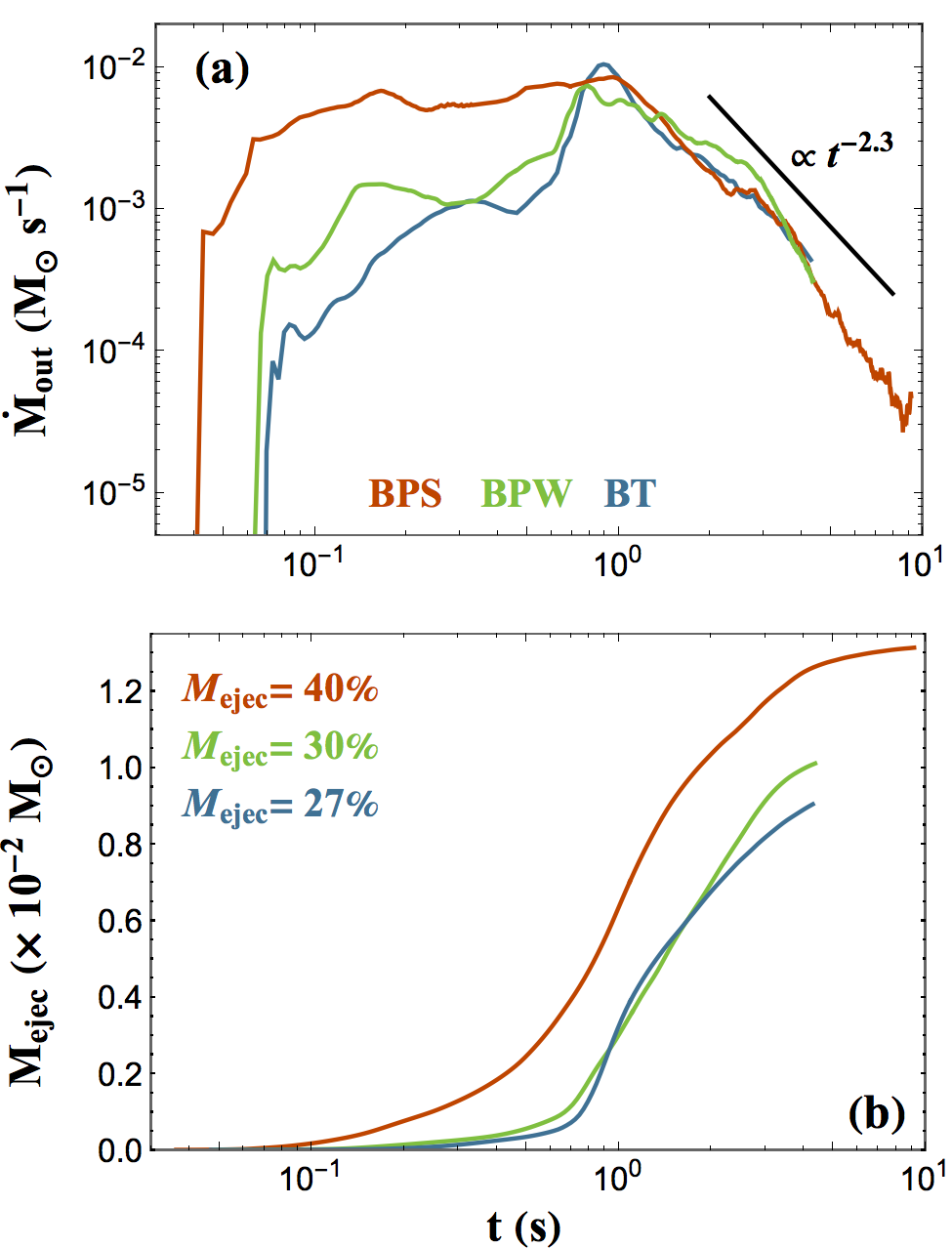}
\caption{The rest mass outflow rate's $\dot{M}_{\rm out}$ (panel a), evaluated at $r_{\rm out} = 10^9 \, {\rm cm} \approx 2000 \, r_g$, strong dependence upon the post-merger field geometry (see legend) demonstrates that strong poloidal flux is required for launching prompt mass outflows. The time in which the initial outflow reaches $r_{\rm out}$ depends on the post-merger geometry, stemming from the developing and saturation time of the MRI as well as the velocity of the outflows' strong dependence on the post-merger field geometry (see Table~\ref{table:model_results}). The large variation in $\dot{M}_{\rm out}$ for times $\lesssim 1$~s presents itself as a $\sim 12 \%$ ($\sim 4 \times 10^{-3} \, M_\odot$) difference in the amount of ejected material $M_{\rm ejec}$ (panel b). At late times ($t \gtrsim 1$~s), $\dot{M}_{\rm out}$ becomes insensitive to the post-merger geometry, displaying a temporal trend of $\propto t^{-2.3}$ in all models, resulting in a flattening of $M_{\rm ejec}$. A coloured version of this plot is available online.}
\label{fig:m_dot_out}
\end{figure}

Why has no simulation seen the development of strong BH magnetic fields in simulations of compact discs with purely toroidal initial magnetic field? There are several possible explanations. First, BH magnetic flux becomes substantial (Fig.~\ref{fig:phi_BH_BT}) and the jets become noticeably strong relative to the accretion flow (Fig.~\ref{fig:e_dot_em}) only at $t\gtrsim 3\;\text{s}$. This corresponds to an extremely long duration of the simulation in terms of BH light crossing times, $t\sim 2\times10^5 \, r_g/c$, much longer than the typical simulation duration of $\sim 10^4 \, r_g/c$. Thus, previous simulations might not have been long enough to observe this effect. Second, in order to see the dynamo action, we needed to use very high resolutions, $512\times256\times128$ cells (see Table~\ref{table:model_resolutions}). We found that while a simulation at twice as small resolution (i.e.$256\times128\times64$ cells) marginally resolved the toroidal MRI, it did not resolve the poloidal MRI, and did not show noticeable signs of large-scale poloidal magnetic flux dynamo.

We note that in the context of radially-extended accretion discs, \citet{liska_dynamo_2018} found the operation of large-scale poloidal flux dynamo and the formation of strong jets, with $\eta_{\rm jet} \gtrsim 1$. This is comparable to the jet efficiency we find, but only at very late times. In fact, it takes our simulations three times longer than those of \citet{liska_dynamo_2018} to reach $\eta_{\rm jet}\sim 1$. Why is this so? If the large radial extent of the disc is a prerequisite for the dynamo to operate efficiently and produce powerful jets, it would take our small disc a substantial amount of time until it appreciably expands radially. Importantly, unlike \citet{liska_dynamo_2018}, our jets also show polarity flips. 
We discuss potential reasons for this difference in Sec.~\ref{sec:tor-magn-geom}.

\subsection{Mass Outflows}
\label{sec:mass-outflows}

Mass outflows and their composition are particularly important as they determine the luminosity, color, and duration of the kilonova (see Secs.~\ref{sec:properties} and~\ref{sec:kilonova}). We quantify the ejecta by measuring the mass outflow rate \mdotout{} through a sphere of radius $r_{\rm out} = 10^9 \, {\rm cm} \approx 2000 \, r_g$. This is sufficiently far from the BH to avoid the interactions with the turbulent and ``viscously'' expanding accretion disc.\footnote{The disc eventually does expands out to such large radii, however, by that time it has very low density and carries little mass. } 
As shown in Fig.~\ref{fig:m_dot_out}(a), the outflows reach $r_{\rm out}$ earliest for strong post-merger poloidal magnetic fields, model BPS, followed at later times by weak poloidal, BPW, and purely toroidal, BT, models. This time difference results from not only the MRI reaching saturation earlier for stronger poloidal magnetic fields, but also from stronger poloidal fields launching faster outflows, as seen in Table~\ref{table:model_results}. Namely, the average radial velocity\footnote{The average radial velocities, determined by $\langle v_r \rangle = \oiint ((\rho+u+P)u^r u^r + P g^{rr})\, {\rm d}A / \oiint \rho \, u^r \, {\rm d}A$, are slightly higher than those found in \citetalias{fernandez2018} due to averaging over momentum rather than density.} of the ejecta for the BPS model is $\langle v_r \rangle \sim 0.18 \, c$, much higher than $\sim 0.08 \, c$ for BPW and $\sim 0.05 \, c$ for BT.

The amount of ejected material also varies by model, as shown in Fig.~\ref{fig:m_dot_out}(b). For instance, in the BPS model, the mass outflow rate rapidly ramps up in a fraction of a second and plateaus at $\mdotout \sim 10^{-2} M_\odot\,{\rm s}^{-1}$. In contrast, in weak poloidal, BPW, and toroidal, BT, models the outflows remain about an order of magnitude weaker and catch up to the BPS model only by the end of the first second. This implies that strong post-merger poloidal magnetic flux is conducive to launching prompt mass outflows.
Interestingly, mass outflows are much more similar at late times, past the first second: the outflow rate in all 3 models largely decays as a power-law, $\mdotout\propto t^{-2.3}$, suggesting that qualitatively the ability of post-merger systems to launch outflows at late times becomes insensitive to the post-merger magnetic field geometry.  The above differences, primarily the prompt mass ejection, lead in the strong poloidal field BPS model to an overall largest mass ejection, carrying $40\%$ of the initial torus mass ($0.013 \, M_\odot$). This is a third more than the $\sim 30 \%$ ejected fraction ($0.01 \, M_\odot$) for BPW and $\sim 27 \%$ ($0.009 \, M_\odot$) for BT models. 

\begin{figure}
\centering
\includegraphics[height=0.32\textwidth]{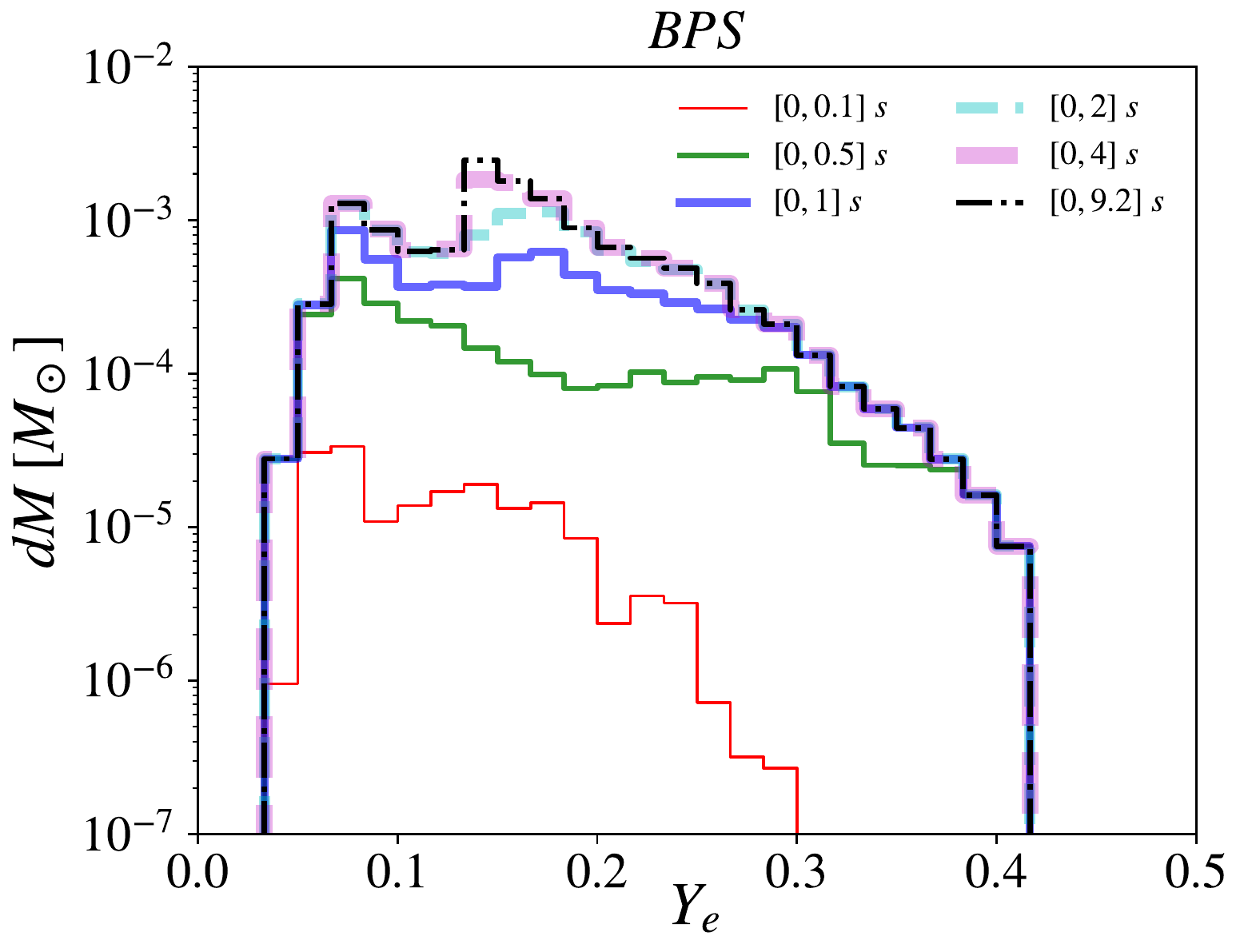}
\includegraphics[height=0.32\textwidth,trim=0.cm 0cm 0cm 0, clip]{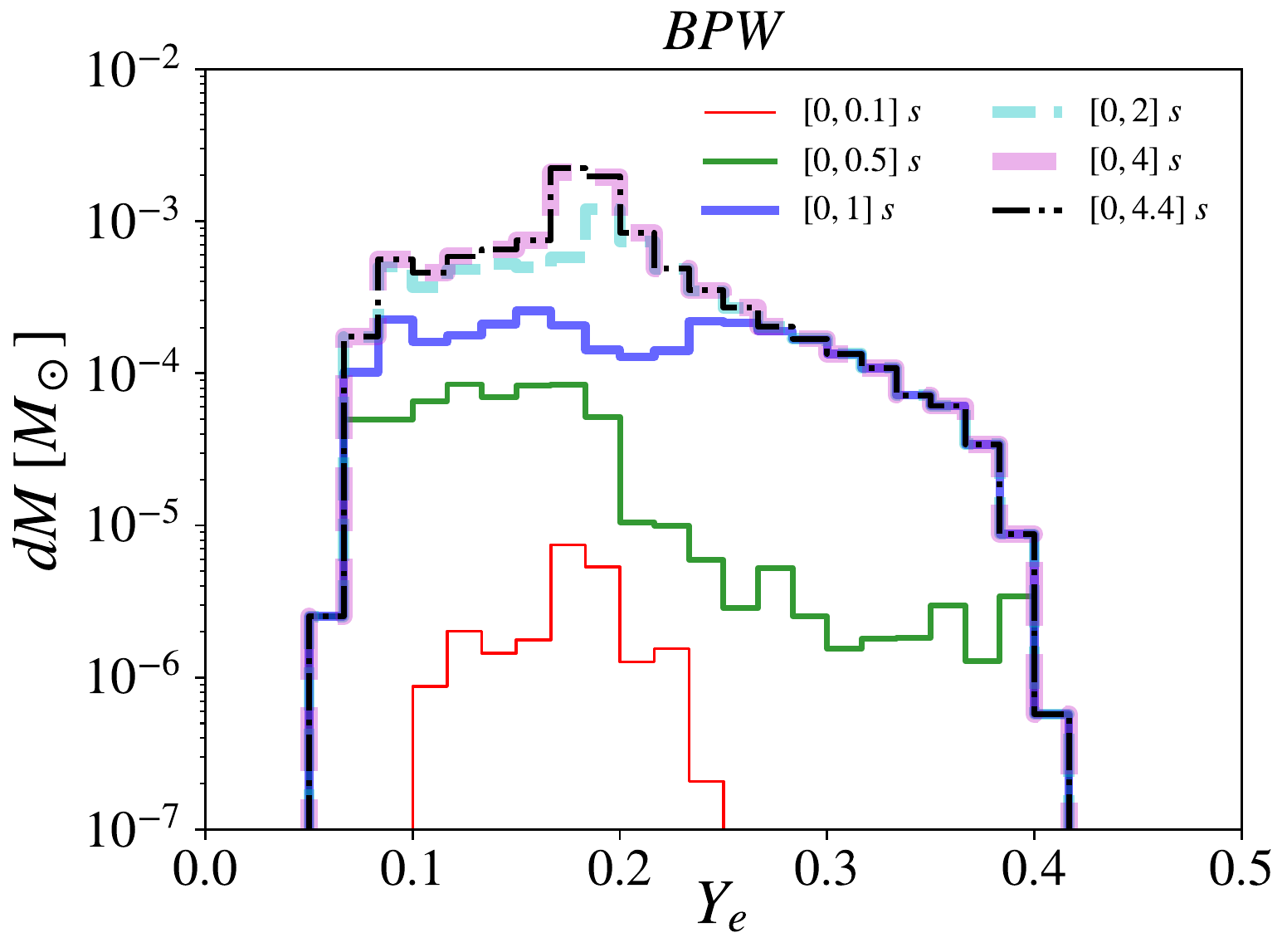}
\includegraphics[height=0.32\textwidth,trim=0.cm 0cm 0cm 0, clip]{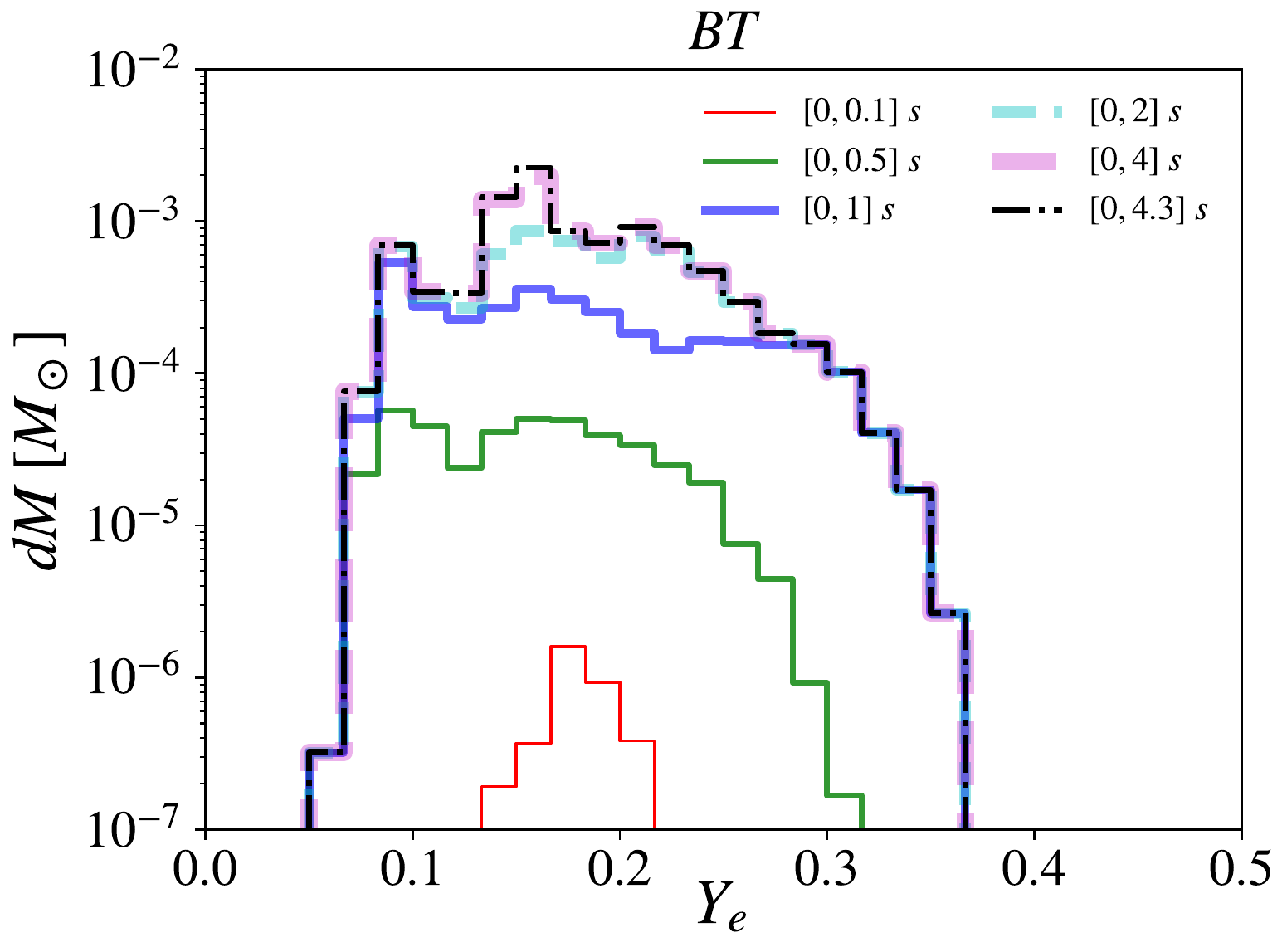}
\caption{Histograms of the cumulative amount of ejected material passing through $r_{\rm out} = 10^9 \, {\rm cm} \approx 2000 \, r_g$ (in addition, see Fig.~\ref{fig:m_dot_out}) vs electron fraction $Y_{\rm e}$, for each post-merger geometry. Increasing the post-merger poloidal field strength not only ejects more material at earlier times, but also spreads this material over a broader range of $Y_{\rm e}$ values, producing a more extended lanthanide-poor (i.e. $Y_{\rm e} \geq 0.25$) region, influenced by the increasing importance of positron capture. A colour version of this plot is available online.}
\label{fig:histograms_dm_dYe}
\end{figure}

\begin{figure}
\centering
\includegraphics[height=0.475\textwidth]{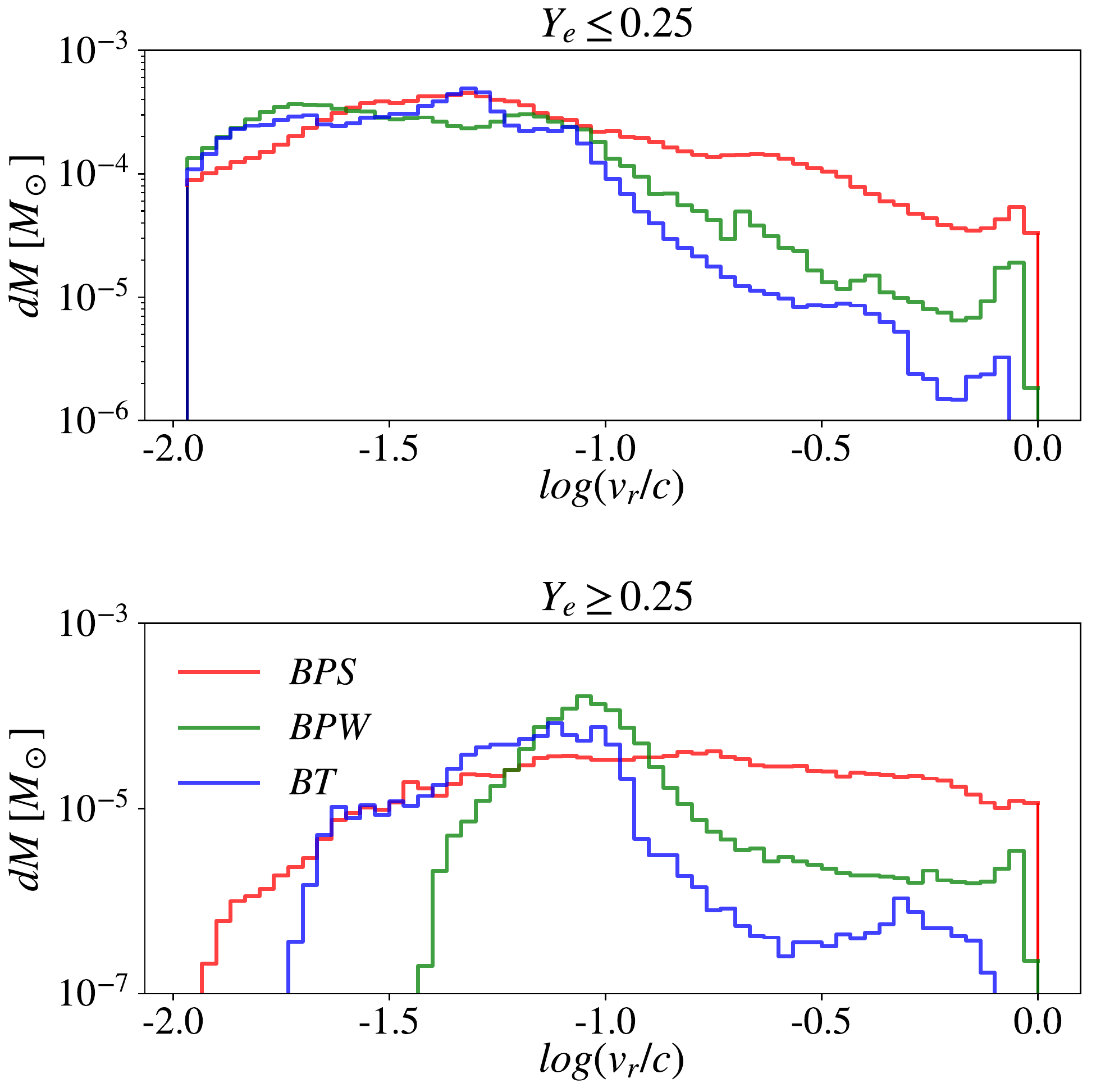}
\caption{Histograms of the cumulative amount of ejected material passing through $r_{\rm out}$ vs the logarithm of the radial velocity (normalized to $c$ and equally spaced in $60$~bins ranging from $-2$ to $0$) of the lanthanide-rich ($Y_e \leq 0.25$, top panel) and lanthanide-poor ($Y_e > 0.25$, bottom panel) regions (see also Fig.~\ref{fig:m_dot_out}). Low $Y_e$-material is characterized by slower radial velocities which are spread over a large range of obtainable $v_r$ values whereas low $Y_e$-material has faster velocities confined within a narrower range of $v_r$. As displayed in Fig.~\ref{fig:m_dot_out}, weaker or more toroidal post-merger field geometries eject less material and at lower velocities. A coloured version of this plot is available online. }
\label{fig:v_r_out}
\end{figure}

To isolate the effects of post-merger magnetic fields, \citetalias{fernandez2018} compared the strong poloidal field BPS model to an otherwise identical hydrodynamic model. They found that the strong poloidal magnetic fields in the BPS model ejected about twice as much mass as in the hydrodynamic model, primarily because the hydrodynamic model was missing the mass ejection during the first second after the merger.
Even our torodial BT model ejects more mass (about a third more) than the hydrodynamic models of \citetalias{fernandez2018}.

\begin{figure*}
\centering
\includegraphics[height=0.6\textwidth]{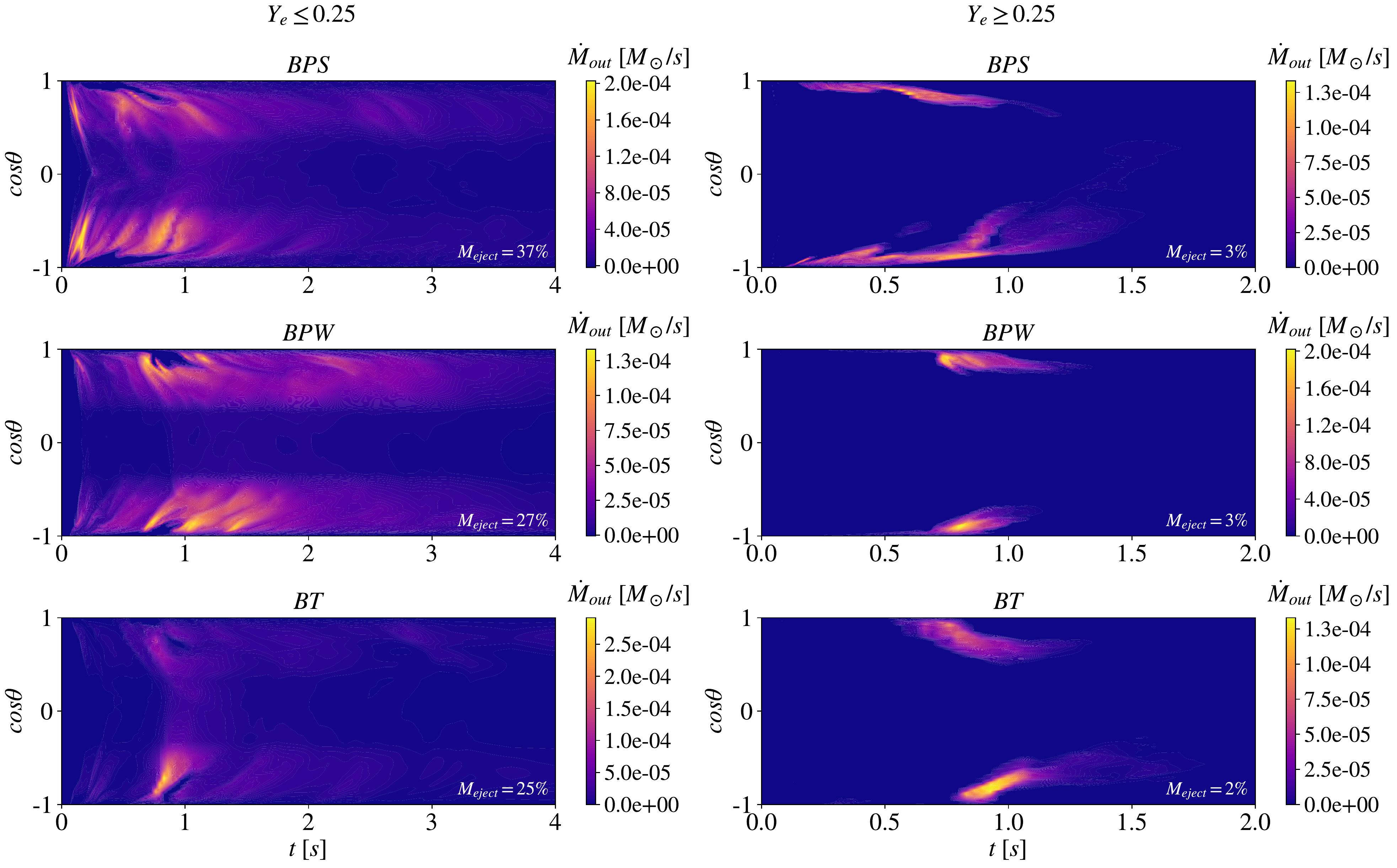}
\caption{A space-time diagram of $\dot{M}_{\rm out}$ (integrated over the $\phi$-direction), as seen on a sphere of radius $r_{\rm out} = 10^9 \, {\rm cm} \approx 2000 \, r_g$ as a function of $\cos\theta$ and time $t$ for all post-merger field geometries (please see the colour bars). The three rows, from top to bottom, show the results for the BPS, BPW, and BT models, respectively. The left panels show lanthanide-rich material ($Y_{\rm e} \leq 0.25$), while the right panels show lanthanide-poor material ($Y_{\rm e} \geq 0.25$). The time range focuses on the period containing most of the ejecta. In the bottom right corner of every panel the total ejected mass is shown as the percentage of the initial torus mass $0.033 \, M_\odot$. We find that low-$Y_e$ material ($Y_e \leq 0.25$) is ejected in large polar regions, extending from the equatorial plane of the disc to regions close to the jet. Higher-$Y_e$ material ($Y_e \geq 0.25$), however, is ejected in much narrower polar regions near the jet. For 3D visualizations, see the temporal snapshots in Fig.~\ref{fig:kilonova_snapshots} and the supplementary videos (\url{https://goo.gl/ct7Htx}). A coloured version of this plot is available online. }
\label{fig:dmdt_rout_ye_separated}
\end{figure*}

\subsection{Outflow Composition}
\label{sec:properties}

In Secs.~\ref{sec:energy-outflows} and~\ref{sec:mass-outflows}, we analyzed the energetics and mass of our outflows. However, these outflows are expected to consist of material with a range of compositions with spatially varying $Y_{\rm e}$ values. Fig.~\ref{fig:histograms_dm_dYe} shows, at different times, the breakdown of ejecta mass $M_{\rm out}$ into bins of $Y_{\rm e}$. Because the post-merger torus is initialized with an electron fraction of $Y_{\rm e} = 0.1$, it is not surprising that at early times $\lesssim 0.1$~s the mass composition is lanthanide-rich\footnote{Nuclear reaction network calculations show that this critical value of $Y_{\rm e} \sim 0.25$ separates the point at which no lanthanides are formed \citep{Lippuner2015}. These elements are key for the opacity and hence, the color of the kilonova \citep{kasen2013}.} (i.e $Y_{\rm e} \leq 0.25$) and centered around $Y_{\rm e} \sim 0.1$. However, the amount of material centered at $Y_{\rm e} \sim 0.1$ strongly depends upon the post-merger geometry, with the BPS model containing more mass than the weaker field BPW and BT models. This trend can also be explained by a similar argument provided in Sec.~\ref{sec:mass-outflows} (additionally, see Fig.~\ref{fig:m_dot_out}), namely stronger poloidal fields launch stronger and faster outflows thereby expelling more mass. For times in between $\sim 0.1 - 1$~s, there is an increase in the amount of material passing through $r_{\rm out}$ for all post-merger geometries with a spread in $Y_{\rm e}$ extending to larger values $\gtrsim 0.25$. These regions form close to the BH at times $t \lesssim 0.3$~s when positron capture becomes increasingly important due to the increasing entropy in the disc (i.e. at fixed radius, density decreases while the temperature remains roughly constant), which results in a higher abundance of positrons and hence more capture, effectively increasing $Y_{\rm e}$. 
We find that the mean electron fraction within all ejecta is almost independent of the post-merger geometry, with $\langle Y_{\rm e}\rangle$ being $0.16$ for BPS, $0.19$ for BPW, and $0.18$ for BT (see Table~\ref{table:model_results}).

As the interpretation of the lanthanide-rich and poor regions has direct applicability to the observed kilonova (see Sec.~\ref{sec:kilonova} for details), it is helpful to investigate the physical and geometrical properties of the two regions. In Fig.~\ref{fig:dmdt_rout_ye_separated}, we show how the mass outflow rate $\dot{M}_{\rm out}$, averaged over the $\phi$-direction, of the lanthanide-rich and poor matter is spread over time $t$ and angle $\theta$. 
Within the lanthanide-rich regions, the amount of material traversing through $r_{\rm out}$ follows a similar trend presented in Figs.~\ref{fig:m_dot_out} and~\ref{fig:histograms_dm_dYe}, with a majority of the material reaching $r_{\rm out}$ at earlier times in the BPS model and later for the BPW and BT models. For all geometries, the lanthanide-rich material is concentrated in the regions between the relativistic jet and the equatorial plane, with the peak amount of material being associated with the peak in $\dot{M}_{\rm out}$ seen in Fig.~\ref{fig:m_dot_out}(a). 
Although the post-merger field geometry governs the total amount of ejected material, it weakly influences the ejected fraction (normalized to the initial torus mass ) of lanthanide-rich gas, $\sim 90 \%$ of the total ejected material (see Table~\ref{table:model_results}).
The material passes through $r_{\rm out}$ with mildly relativistic speeds; namely $v_r \sim 0.01 - 0.1 \, c$, as displayed in the top panel of Fig.~\ref{fig:v_r_out}. The time averaged radial velocity of the lanthanide-rich regions is $\langle v_r \rangle_{\rm red} \sim 0.17  \, c$ for the BPS model, much faster than $\sim 0.07  \, c$ for the BPW model and $\sim 0.05  \, c$ for the BT model. Even larger speeds are found in material moving within the jet, however, it is orders of magnitude less dense than the surrounding winds (see Fig.~\ref{fig:B_phi_temporal_snapshots} for comparison).

\begin{figure*}
\centering
\includegraphics[height=0.3\textwidth,trim=5.8cm 5cm 5.8cm 3.5cm, clip]{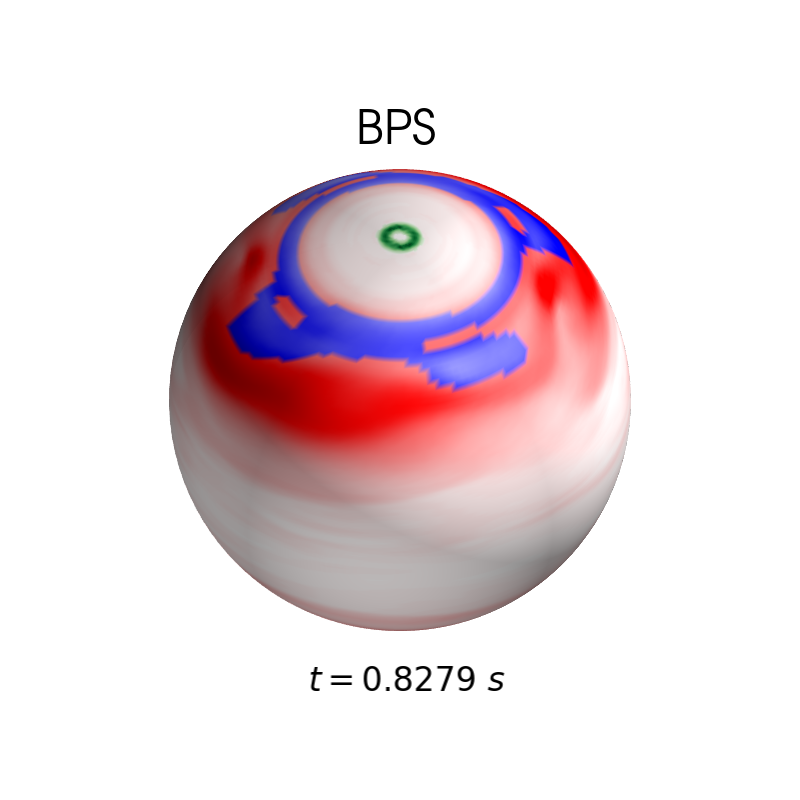}
\hspace{1cm}
\includegraphics[height=0.3\textwidth,trim=5.8cm 5cm 5.8cm 3.5cm, clip]{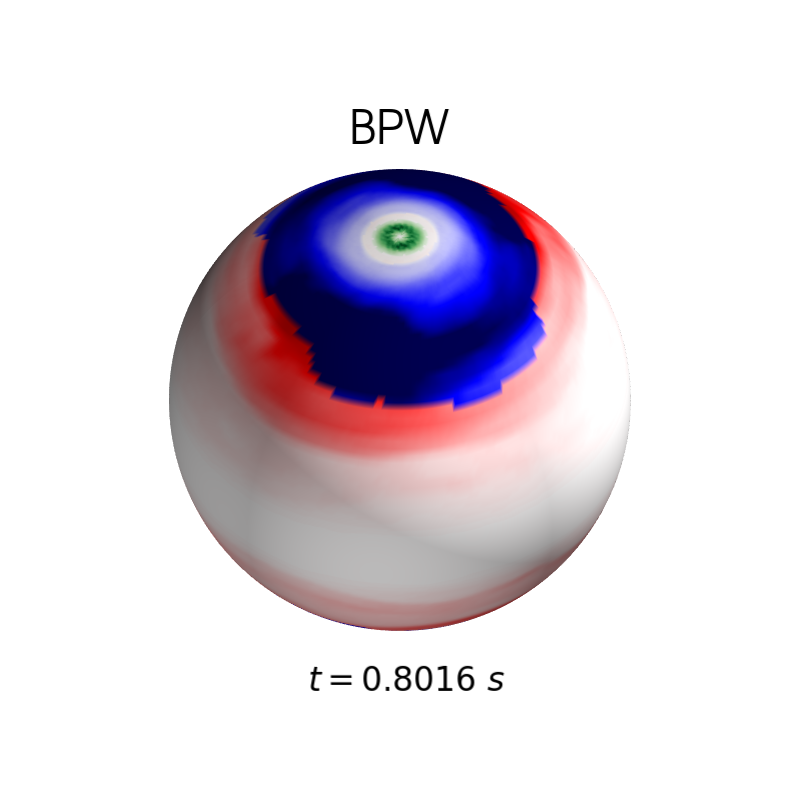}
\hspace{1cm}
\includegraphics[height=0.3\textwidth,trim=5.8cm 5cm 5.8cm 3.5cm, clip]{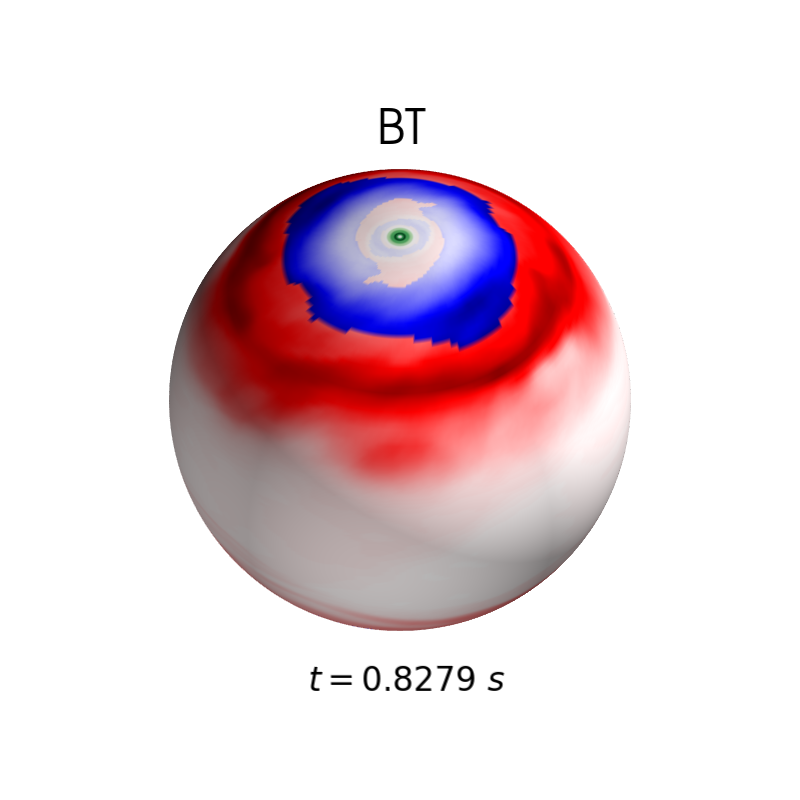}
\caption{Temporal snapshots ($t \approx 0.8$~s) of the kilonova and jet components painted over a uniform sphere of radius $r_{\rm out} = 10^9 \, {\rm cm} \approx 2000 r_g$. The color intensity of the blue and red kilonova components represent the mass-weighted rest-mass outflow rate $\dot{M}_{\rm out}$ (see Figs.~\ref{fig:m_dot_out} and~\ref{fig:dmdt_rout_ye_separated}) while the jet (green) intensity is weighted by its power (see Fig.~\ref{fig:e_dot_em}). The blue component is confined within a much narrower region (polar width $\Delta \theta \sim 15^\circ{-}20^\circ$) than the red component. Initially obscured by the red component, the blue component moves faster, punching through and eventually overtakes the red component. The relativistic jet within each model is tightly collimated by the surrounding disc winds, obtaining opening angles $\theta_{\rm jet} \lesssim 15^\circ$ (see also Fig.~\ref{fig:theta_jet}).  For videos displaying the full time evolution, see \url{https://goo.gl/ct7Htx}.  A coloured version of this plot is available online. }
\label{fig:kilonova_snapshots}
\end{figure*}

For the lanthanide-poor regions, the material begins to pass through $r_{\rm out}$ at $\sim 0.1$~s for the BPS model and $\sim 0.5$~s for the weaker field BPW and BT models and continues to pass through for up to $\sim 1$~s in all models, until positron capture becomes less important. The amount of ejected material contained within the lanthanide-poor region is $\sim 3 \%$ (i.e. $\sim 10^{-3} \, M_\odot$) within all models, independent of the post-merger geometry (see Table~\ref{table:model_results}). The material passing through $r_{\rm out}$ is contained within a much narrower angular region than the lanthanide-rich material, with polar width of $\Delta \theta \sim 15^\circ - 25^\circ$. We illustrate this in Fig.~\ref{fig:kilonova_snapshots}, by plotting on a sphere of radius $r_{\rm out}$ for each post-merger geometry a temporal snapshot (at $t\sim 0.8$~s) of the mass-weighted lanthanide-rich (red) and poor (blue) regions\footnote{For videos displaying the full time evolution of Fig.~\ref{fig:kilonova_snapshots}, see: \url{https://goo.gl/ct7Htx}.}, in addition to the relativistic jets (green). 
At early times, when the lanthanide-poor ejecta initially crosses $r_{\rm out}$, it is concentrated near the poles. For later times, the lanthanide-poor gas appears to emerge at larger polar angles (closer to the equatorial plane). This gas, initially obscured\footnote{This obscuration occurs very early in the disc evolution and should not be mistaken with obscuration occurring within the late-time (i.e. $t \gg 1$~s) kilonova light curve.} by the lanthanide-rich material, passes through $r_{\rm out}$ at relativistic speeds larger than the lanthanide-rich material, namely $v_r \gtrsim 0.03 \, c$ with velocities reaching up to $c$, as presented in the bottom panel of Fig.~\ref{fig:v_r_out}. The time averaged radial velocity of the lanthanide-poor material is $\langle v_r \rangle_{\rm blue} \sim 0.3  \, c$ for the BPS model, $\sim 0.16  \, c$ for the BPW model, and $\sim 0.08  \, c$ for the BT model, eventually punching through and overtaking the lanthanide-rich material. This ejection is mildly asymmetric relative to the equatorial plane for the BPS and BT models, with more material being ejected through the southern hemisphere. This anisotropy manifests itself as a $\sim 3\times 10^{-4} \, M_\odot$ difference in the lanthanide-poor region of our BPS model and a $\sim 1.3\times 10^{-3} \, M_\odot$ difference in the lathanide-rich region of the BT model.

\section{Discussion}
\label{sec:discussion}

\subsection{Poloidal Magnetic Geometry and Dynamically-Important Magnetic Fields}
\label{sec:jet-energetics}

We have found that for a range of magnetic field geometries, post-merger accretion discs can naturally develop dynamically-important BH magnetic flux that turns them into MADs, providing us with an inside view of the mass accretion rate on the BH from the jet power. Namely, at early times, the jet power is set by the amount of large-scale poloidal magnetic flux present in the accretion flow, whereas at late times it is set by the mass accretion rate. This outcome is insensitive to the post-merger magnetic field geometry which suggests that it is a robust phenomenon, even though it has not been found previously.
Until now, smaller torii extending out to $\lesssim 50 \, r_g$ embedded with a single poloidal field loop, have been found to lead to Standard And Normal Evolution \citep[SANE,][]{narayan2012} discs, in which the gas pressure dominates the magnetic pressure in the disc, \citep[e.g.,][]{gammie2003,2005ApJ...630L...5M,hawley2006}. Indeed to obtain MADs, researchers have previously opted for very large accretion discs, much larger than those typically expected in the context of compact binary mergers: the large size of the discs allowed them to contain a large enough poloidal magnetic flux to flood the BH \citep{2015ASSL..414...45T,2015SSRv..191..441H}. For instance, to obtain MADs, \citet{Tchekhovskoy2011} considered discs of a large extent $\sim 5\times 10^4 \, r_g$ while \citet{McKinney2012} simulated even larger, unbound discs that extended out to infinity. \citet{narayan2012} obtained similar results while focusing on discs that extended out to $10^3 \, r_g$.

While such large discs are naturally expected in active galactic nuclei (AGN), binary mergers lead to much smaller discs. How can a small disc in a binary merger turn MAD?
Instead of starting with a large amount of magnetic flux, a disc can start with little flux and evolve to the point when very little of the initial gas is left. At this late time, what was initially a weak and subdominant magnetic flux can become a dynamically-important one. In fact, this is a natural way of producing MADs in any system whose mass accretion rate decreases over time, such as tidal disruption events \citep{2014MNRAS.437.2744T} and core-collapse gamma-ray bursts \citep{Tchekhovskoy2015}. In this work, we demonstrated that MADs can naturally develop in 3D numerical simulations of initially small accretion discs, as typical for binary mergers (see \citealt{Proga2006} for a 2D analog). However, in order for a MAD state to occur, an unusually long evolution time (by GRMHD simulation duration standards) is required, which might explain why this effect has not been previously seen.

In our strong poloidal magnetic field simulation, model BPS, the magnetic fields became dynamically-important around $t_{\rm MAD} \approx 0.5$ seconds after the merger, when the dimensionless BH magnetic flux reaches a critical value, $\phi_{\rm BH}=\phi_{\rm MAD} \simeq 50$ (see Fig.~\ref{fig:phi_BH}(b)). In a weaker poloidal magnetic field model, BPW, this happens at a few times later time, $t_{\rm MAD}\approx 2$~s. This makes sense, since the stronger the initial magnetic flux, the earlier it becomes dynamically important relative to the decreasing pressure of the accretion disc.

Interestingly, Fig.~\ref{fig:phi_BH}(b) shows that even in complete absence of any post-merger poloidal magnetic flux, as in our model BT, the dimensionless magnetic flux reaches $\phi_{\rm BH} \approx 35$, more than half way to the critical $\phi_{\rm MAD}$ value, by the end of the simulation. This suggests that given a longer duration (e.g. $\sim10$~s), the BH magnetic flux can become dynamically-important even for this purely toroidal post-merger magnetic flux geometry. We discuss how this can happen in Sec.~\ref{sec:tor-magn-geom}.

\subsection{Toroidal Magnetic Geometry and Striped Jets}
\label{sec:tor-magn-geom}

Poloidal magnetic flux is a crucial prerequisite for jet formation.
Indeed, it is the winding of the poloidal magnetic field by the BH \citep{blanford1977} or the inner regions of the disc \citep{1982MNRAS.199..883B} that is typically associated with magnetically-powered outflows. However, the shear between two merging neutron stars is expected to amplify the toroidal magnetic field component, naturally leading to a toroidally-dominated field geometry (the field direction might undergo polarity flips on small scales due to the Kelvin-Helmholtz instability). In the absence of sufficiently strong poloidal magnetic fields, how do binary mergers manage to produce jets at all?

Our toroidal field simulation, model BT, might shed light on this long-standing problem. We find that as the simulation progresses, the accretion flow spontaneously develops poloidal magnetic field loops above and below the equatorial plane (see the movies in the Supplementary Information and at \href{https://goo.gl/ct7Htx}{https://goo.gl/ct7Htx}). This behaviour is similar to that of an $\alpha{-}\Omega$ poloidal flux dynamo \citep{moffatt1978}, in which buoyancy and Coriolis forces work together to convert toroidal flux into poloidal magnetic flux loops. Because of the chaotic nature of the dynamo, the magnetic flux polarity varies randomly from one loop to another. Once the newly formed magnetic flux loops reach the BH, they power jets of alternating magnetic flux polarity, or striped jets.
At large distances in the jet, magnetic reconnection in the current sheets separating the regions of opposite polarity can provide natural dissipation sites responsible for high-energy GRB jet emission \citep{giannios2018}.
Note that the total poloidal flux on the BH event horizon, determined by eqn.~\eqref{eqn:B_phi_BH}, is greater than the sum of the magnetic fluxes from its individual northern and southern hemispheric components. 
This non-zero difference between $\Phi_{\rm BH}$ and $|\Phi_{\rm BH, north}| + |\Phi_{\rm BH, south}|$ emerges due to the presence of the current sheets near the BH horizon.

Recently, \citet{liska_dynamo_2018} found that radially-extended accretion discs initially threaded with a purely toroidal magnetic field can produce powerful jets. Similar to our work, they see the signs of $\alpha{-}\Omega$ dynamo and the formation in the accretion disc of poloidal field loops of alternating polarity. However, they find that most of the loops become ejected in an outflow, and a single magnetic loop ends up dominating the jet energetics and magnetic flux polarity. Why are our results different than theirs? There can be two possible factors that can suppress outflows in our work and encourage the disc to retain the poloidal flux loops instead of ejecting them.
First, our smaller disc is more tightly bound and is therefore less conducive to outflows. Second, our neutrino cooling makes the disc even more tightly bound, resulting in additional outflow suppression. Both of these effects might encourage the disc to retain most of the freshly generated poloidal magnetic flux loops and encourage them to accrete on the BH, leading to alternating BH magnetic flux and striped jets. In future work, we will investigate the robustness of this phenomenon and its relationship to the size of the accretion disc, the presence of radiative or neutrino cooling, and the sensitivity of the results to the numerical resolution.

\subsection{Jet Opening Angles \& Isotropic Equivalent Energy}
\label{sec:jet-opening-angles}

\begin{figure}
\centering
\includegraphics[height=0.285\textwidth]{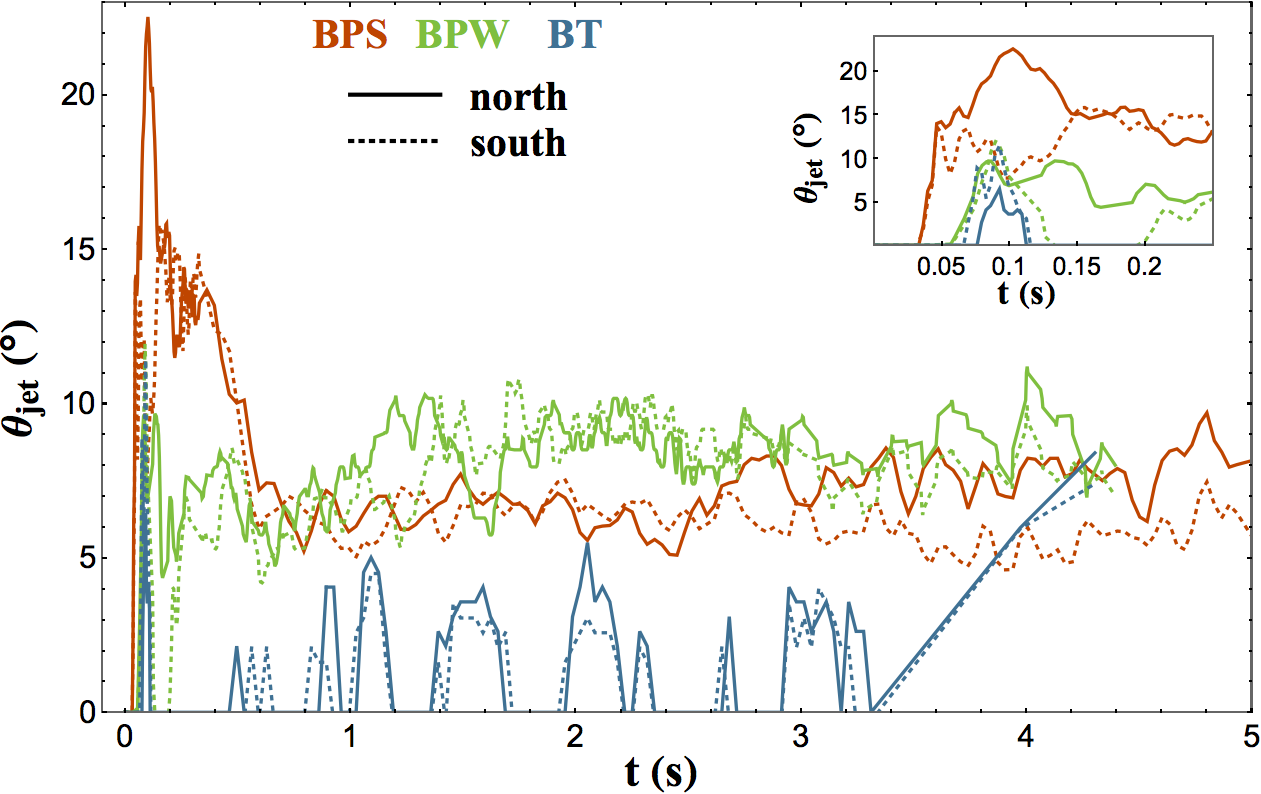}
\caption{The opening angles of both jets (see the legend for explanation of line types) substantially differ between purely toroidal and purely poloidal post-merger magnetic geometries (see legend), with the former, model BT, having smaller opening angles. In the BT model, the surrounding disc winds disrupt the jet (i.e. $\theta_{\rm jet} = 0^\circ$, see also Fig.~\ref{fig:e_dot_em}), forcing the jet to punch through the material, producing a tightly collimated angle. The range of opening angles in our simulations, $5{-}20^\circ$ is roughly consistent with the range of inferred opening angles in short GRBs \citep{Fong2015}. A coloured version of this plot is available online.}
\label{fig:theta_jet}
\end{figure}

X-ray and optical afterglow observations of GRBs exhibiting a steepening in their temporal decline are often characterized by jet breaks \citep{Soderberg2006,Nicuesa2011,Fong2012b,Fong2014}. The time associated with these breaks can uncover characteristic properties of the jet, such as the jet opening angle $\theta_{\rm jet}$ \citep{sari1999,Frail2001}. Although only a few short GRBs exhibit such a break,  \citet{Fong2015} were able to estimate their opening angles to span a wide range from $\sim5$ to a few tens of degrees, with a median jet opening angle of all measured short GRBs, $\langle \theta_{\rm jet} \rangle \approx 16^\circ \pm 10^\circ$.  

These observations raise an important question: what is the jet collimating agent in short GRBs? In long-duration GRBs, which can also be tightly collimated \citep[into opening angles as small as a few degrees,][]{Frail2001,2010ApJ...711..641C}, a natural collimating agent is the radially-extended stellar envelope of size $\gtrsim 10^4 r_g$ \citep{2010NewA...15..749T}. It does not appear plausible, however, that a compact post-merger remnant disc extending out to $\sim50r_g$ can manage to collimate short GRB jets into the smallest observed short GRB opening angles of $\sim5^\circ$. What is the way out of this conundrum? Our simulations reveal that magnetized turbulence leads to angular momentum transport and viscous-like spreading of the disc from the initial size of $\sim50r_g$ to $\gtrsim 10^3r_g$. While this is an order of magnitude smaller than the size of the stellar envelope in long GRBs, outflows launched from such an extended disc could substantially collimate the jets. Does this lead to sufficiently small jet opening angles, $\theta_{\rm jet}$, that span the range of short GRB observations?

\begin{figure}
\centering
\includegraphics[height=0.86\textwidth]{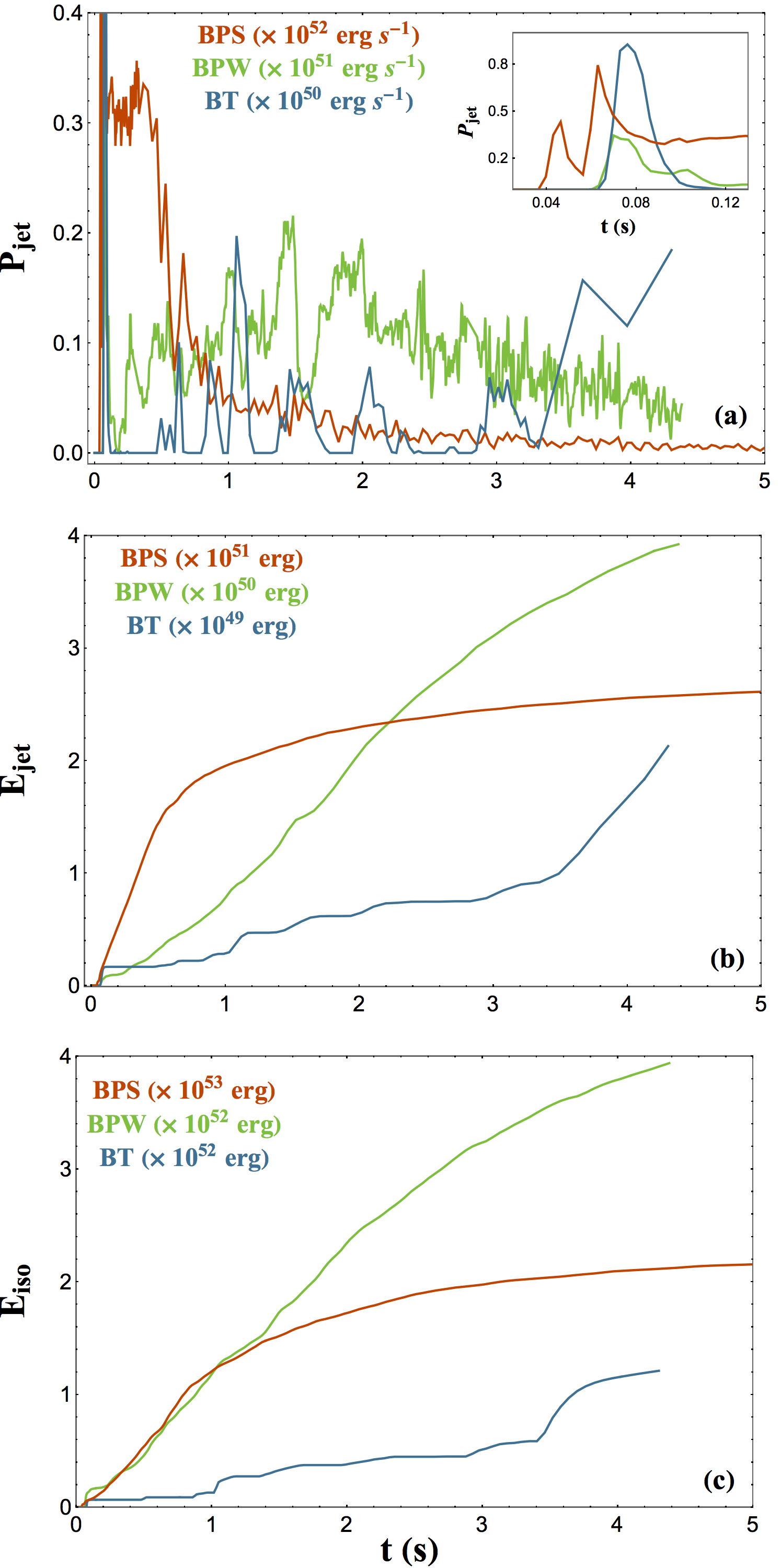}
\caption{Not only does the post-merger field geometry drastically effect the jet power (panel a), but it also results in a large difference in the cumulative jet energy (panel b) and its isotropic equivalent (panel c). For the BPS model, a majority of the energy passes through $r_{\rm out}$ within the first $\sim 1$~s, while for weaker fields passes through by $\sim 4$~s. In the BT model, the intermittency of the jet results in a smaller jet energy. A coloured version of this plot is available online. }
\label{fig:E_jet}
\end{figure}

Figure~\ref{fig:theta_jet} shows that at early times ($t \sim 0.1$~s) for all post-merger geometries, $\theta_{\rm jet}$ reaches its peak, the largest being $\theta_{\rm jet} \sim 24^\circ$ for the BPS model, followed by $\sim 10^\circ$ for the BPW model, and $\sim 7^\circ$ for the BT model. These differences in the opening angle could result from: i) the lack of sufficient material surrounding the jet at early times, required to tightly collimate it, as the jet material launched around the polar axis moves faster than the surrounding disc winds\footnote{We note that in our simulations, we do not consider neutrino/anti-neutrino annihilation. Such effects can deposit enough energy into the polar regions to drive mildly relativistic outflows, clearing the poles of baryons \citep{Fujibayashi2017,Foucart2018}.}, and/or ii) the power contained within the jet is $\gtrsim 50$ times larger for the BPS model than the other models (see Fig.~\ref{fig:e_dot_em}).
For late times, $t\gtrsim 0.5$~s, in our poloidal field models, a large fraction of material reaches $r_{\rm out}$ at larger polar angles, producing a tight collimation of the jets. For BPS and BPW models, the time-average opening angles (averaged over their the jet activity period, i.e. $t \lesssim 1$~s, and also averaged over both jets) is $\langle \theta_{\rm jet} \rangle \sim 13^\circ$ and $\sim 6.4^\circ$, respectively (see Table~\ref{table:model_results}).  
For the purely toroidal BT model, the intermittence of $\theta_{\rm jet}$ follows that of $P_{\rm jet}$ (see Fig.~\ref{fig:e_dot_em}): the surrounding disc winds disrupt the jets leading to $\theta_{\rm jet} = 0^\circ$. As the jets reform near the BH, they have to drill through the disrupted material, resulting in a tighter collimation with a time-average value (averaged over $1$~s) of $\langle \theta_{\rm jet} \rangle \sim 4.6^\circ$.

The investigation of the jet opening angle’s dependence on the post-merger geometry has important implications on the inferred energies of the resulting afterglow. It is difficult to measure directly the intrinsic jet energy, $E_{\rm jet}$. More easily accessible is its isotropic equivalent energy, $E_{\rm iso}$, which is related to $E_{\rm jet}$ by the beaming factor $f_b \equiv 1 - (\cos\theta_{\rm jet, north} + \cos\theta_{\rm jet, south})/2$, such that $E_{\rm iso} = E_{\rm jet}/f_b$. Although the inferred value of $E_{\rm iso}$ from GRB afterglows is model dependent (e.g. assumed particle spectrum, radiative efficiency, density of external ambient medium), a compilation of 38 short GRBs reports a median value of $E_{\rm iso} \sim 3\times10^{51}$~erg while the distribution spreads over a range of $\sim 3 \times 10^{49}$~erg to $\sim 10^{53}$~erg \citep{Fong2015}. Assuming the median jet opening angle reported above \citep[i.e. $\sim 16^\circ$,][]{Fong2015}, this corresponds to a characteristic inferred intrinsic jet energy of $E_{\rm jet} \sim 10^{51}$~erg.  

We show in Figs.~\ref{fig:E_jet}(b) and~(c) the cumulative jet\footnote{The distinction between the jet and disc winds is made by performing a cut on the specific energy flux: $\mu = -T^r_t/(\rho u^r) \geq 2$ (see Sec.~\ref{sec:energy-outflows}).} energy and its isotropic equivalent, as measured at a sphere of radius $r_{\rm out} = 10^9 \, {\rm cm} \approx 2000 \, r_g$ for all three post-merger magnetic geometries. For the BPS model, the large spike in $P_{\rm jet}$ within the first $\sim 1$~s results in $E_{\rm jet}$ quickly reaching an asymptotic value of $\sim 2.5 \times 10^{51}$~erg. At late times $\gtrsim 1$~s, the jet power rapidly drops resulting in a negligible contribution to its energy. Its isotropic equivalent follows a similar trend such that at $\sim 1$~s, a majority of the energy has passed through $r_{\rm out}$, beyond which it slowly reaches an asymptotic value of $E_{\rm iso} \sim 2.2 \times 10^{53}$~erg.
For weaker post-merger poloidal field geometries, such as our BPW model, the jet power remains roughly constant throughout time and does not begin to decrease until $\gtrsim 4$~s. For $E_{\rm jet}$ and $E_{\rm iso}$, this corresponds to a continuous increase in time, eventually leveling off at values of $E_{\rm jet}\sim 3.9 \times 10^{50}$~erg and $E_{\rm iso}\sim 4 \times 10^{52}$~erg, respectively, at $\sim 4$~s. 

For an initially toroidal post-merger geometry, the jets are weak and intermittent (see Figs.~\ref{fig:e_dot_em} and~\ref{fig:E_jet}(a)). Because of this, the jet energy and isotropic equivalent energy are very gradual functions of time, reaching values of $E_{\rm jet}\sim 2 \times 10^{49}$~erg and $E_{\rm iso}\sim 1.3 \times 10^{52}$~erg, respectively. However, it is important to note the late time (i.e. $\gtrsim 3.5$~s) difference between $E_{\rm jet}$ and $E_{\rm iso}$. There is a continuous increase in the former up until the end of the simulation due to the late time increase in the jet power. However, at this time, $\theta_{\rm jet}$ also increases significantly above its time averaged value (see Fig.~\ref{fig:theta_jet}). The increase in both quantities results in a late time decrease in the isotropic equivalent of the jet power presenting itself as a leveling off of $E_{\rm iso}$.

Note that in our simulations the sub-relativistic winds surrounding the jets can carry a fraction of the jet energy in all models: typically, $E_{\rm wind} \sim 5 \times 10^{50}$~erg for BPS, $\sim 10^{50}$~erg for BPW, and $\sim 4 \times 10^{49}$~erg for BT models. This energy will eventually become visible at the forward shock, where the ejecta runs into the ambient medium and produces the afterglow emission. Months-long radio and X-ray afterglow seen from GW170817 is thought to be powered by the relativistic jet \citep{Margutti2018,kathirgamaraju2018,Alexander2018}. 

\subsection{Kilonova Implications}
\label{sec:kilonova}

A detailed analysis of the inferred properties of GW 170817/GRB 170817A has been reported in \citep{kasen2017,Kasliwal2017,Kilpatrick2017}. The total amount of ejected material is estimated within the range of $\sim 0.01 - 0.1$~$M_{\odot}$. 
To match the optical and infrared observations \citep{Arcavi2017,cowperthwaite2017,drout2017}, \citet{kasen2017} modeled the red (i.e. $Y_{\rm e} < 0.25$) and blue (i.e. $Y_{\rm e} > 0.25$) kilonova components with ejected masses $M_{\rm red} \approx 0.04 \, M_{\odot}$ and $M_{\rm blue} \approx 0.025 \, M_{\odot}$ and velocities of $v_{\rm red} \approx 0.1 \, c$ and $v_{\rm blue} \approx 0.3 \, c$, respectively. 
Their analysis suggests that the mechanism for mass ejection is predominantly via outflows from a remnant accretion disc.
Although our simulations are consistent with this interpretation, there are several differences found, for each post-merger field geometry, when making a comparison to observations, as discussed below. 

From our results, we find that the total ejected mass ($0.013 \, M_\odot$ for BPS, $0.01 \, M_\odot$ for BPW, and $0.009 \, M_\odot$ for BT; see Fig.~\ref{fig:m_dot_out} and Table~\ref{table:model_results}) is lower than the values listed above. To obtain values consistent with observational modeling, we would require an initial torus mass $\gtrsim 0.1 \, M_{\odot}$, if we were to simply rescale our results. As shown in the left column of Fig.~\ref{fig:dmdt_rout_ye_separated}, a majority of the ejected mass has $Y_{\rm e} \leq 0.25$.
As compared with observations, all configurations underpredict the ejected mass of the red component by $\lesssim 5$ times. Assuming a simple rescaling of our results to an initial torus mass of $\sim 0.15 \, M_\odot$, the amount of material within the red component (for all post-merger geometries) would be consistent with the inferred values from observational modeling of GW 170817/GRB 170817A. However, the average radial velocity of the red component varies significantly with the post-merger geometry, with the BT model obtaining smaller velocities than what is inferred (see Table~\ref{table:model_results}).

For the blue kilonova component, all three models underpredict the inferred mass by several orders of magnitude ($\sim 10^{-3} \, M_\odot$ from all models as compared to $0.025 \, M_\odot$ from observational modeling). If our results were simply rescaled with the torus mass, we would require $\gtrsim 0.8 \, M_{\rm \odot}$, much larger than what is expected post-merger. Moreover, it is useful to note that there is a difference in the geometrical interpretation of the blue kilonova component. When modeling the emission from the blue component, \citet{kasen2017} took the blue material to be painted over a spherical region within polar angles $0^\circ \leq \theta \leq 45^\circ$. As discussed in Sec.~\ref{sec:properties} (see Figs.~\ref{fig:dmdt_rout_ye_separated} and~\ref{fig:kilonova_snapshots}) for all three models, the blue material at $r_{\rm out} = 10^9 \, {\rm cm} \approx 2000 \, r_g$ is confined within narrower regions of polar width $\Delta \theta \sim 15^\circ - 20^\circ$. It could be, however, that our simulations may not have reached a free-expansion phase. If so, it could be likely that geometry of the blue region would be modified before being observed as a kilonova.

The aforementioned analysis therefore suggest that an initially toroidal field, for our idealized initial conditions, struggles to reproduce the kilonova properties (e.g. mass and velocities) inferred from observational modeling. It should be noted, however, that simply rescaling the mass of the initial torus would not be entirely accurate as discs with larger masses are more opaque to neutrinos, requiring a more elaborate treatment of neutrino cooling than what was used here (\citetalias{fernandez2018}). Moreover for GW 170817, \cite{Shibata2017} estimated the range of potential torus masses to lie within $0.05 - 0.2 \, M_\odot$, putting a constraint on our analysis. In order to provide a more complete model, the inclusion of the post-merger dynamical ejecta and a better neutrino transport scheme would be necessary, which could reduce the tension in the amount of mass ejected within the blue component. 

\section{Conclusions}
\label{sec:summary}

Here, we explored the role of the post-merger magnetic field configuration on the long term disc evolution in the context of NS mergers. 
Beginning with either a purely poloidal or purely toroidal magnetic field within the torus (see Table~\ref{table:model_resolutions} for model initial setup), we find the formation of a relativistic jet whose total power, energy, and opening angle are consistent with typical values inferred from GRBs (see Figs.~\ref{fig:e_dot_em}b,~\ref{fig:theta_jet}, and~\ref{fig:E_jet}).
For all three post-merger magnetic field configurations, we find that the jet power eventually reaches a level of the order of $\dot{M}_{\rm accr} \, c^2$, signifying that the disc has reached a MAD state (see eqn.~\ref{eq:Pjet_mad} and Fig.~\ref{fig:phi_BH}b) and that at late times, the jet power directly tracks the mass accretion rate on the BH. At earlier times, the jet power is roughly constant and not strongly correlated with $\dot{M}_{\rm BH}$, reflecting the large scale poloidal magnetic flux content in the accretion flow.

For the purely toroidal post-merger magnetic field configuration, we find the formation of a jet with energetics consistent with GRBs (Fig.~\ref{fig:E_jet}). 
A dynamo-like process in the accretion disc leads to the formation of alternating magnetic flux, shown in Fig.~\ref{fig:phi_BH_BT}, that powers striped jets. 
If this result holds at higher resolution, the production of current sheets and their reconnection in the jets could power the prompt emission in GRBs \citep[e.g.][]{spruit2001,giannios2006,Beniamini2018}.

Concurrent with the launching of a jet, mass outflows are expelled as disc winds. The driving mechanism of the winds is not fully understood but is most likely a combination of thermal and magnetic effects as well as small contributions from $\alpha$-particle recombination \citep{metzger2008,siegel2018}.  
The total amount of ejected material contained within winds (see Fig.~\ref{fig:m_dot_out}) is found to be smaller than what is inferred from observational modeling of GW 170817/GRB 170817A. However, our simulations assumed an initial torus mass of $0.033 \, M_\odot$.
The initial torus mass, determined by the masses of the binary components and the assumed equation of state, is expected to be larger than what was assumed here \citep{Shibata2017}. 

It is important to note several limitations of our analysis. The first is our choice for the initial electron fraction of $Y_{\rm e} = 0.1$ prescribed within the initialized torus. This value is lower than what is found in typical merger simulations of binary neutron stars \citep{Foucart2015_NSNS,Foucart2016_NSNS,Sekiguchi2016_NSNS} and BH/NS systems \citep{Foucart2015_NSBH,Foucart2017_NSBH}, which have reported an electron fraction ranging from $\sim 0.1 - 0.2$. The second is the lack of neutrino absorption and transport within our models. Recent results of GRMHD simulations combined with a full Monte Carlo neutrino transport method have shown that $\sim 20 \%$ of the early time outflows are blue \citep{Miller2019}. As such, these approximations can lead to underestimating $Y_{\rm e}$ at early times. Therefore, the fraction of the lanthanide-poor ejecta presented here (see Table~\ref{table:model_results}) can be considered as a lower bound.

Our choice for the initial torus mass of $0.033 \, M_\odot$ was made such that the torus remains optically thin and our approximations for neutrino cooling are reasonable. Moreover, our simulations results were completed before the observations of GW170817/GRB170817A. 
The results of using a larger torus, and its comparison with the observed kilonova are currently being explored and will be presented elsewhere. We note, however, that for more massive discs, a more complete treatment of neutrino transport will be necessary \cite[as in, e.g. ][]{Miller2019}.

Although the total amount of material contained within the winds varies with the initial magnetic field configuration, shown in Fig.~\ref{fig:m_dot_out}, its composition remains roughly fixed, with $\sim 90 \%$ being lanthanide-rich material and the remaining $\sim 10 \%$ being lanthanide-poor (see Figs.~\ref{fig:histograms_dm_dYe} and~\ref{fig:dmdt_rout_ye_separated}). A simple rescaling of our results to an initial torus mass of $0.15M_\odot$ leads to lanthanide-rich ejection consistent with the red kilonova component inferred for GW170817, and with mass ejection falling short by a factor of $\sim 6$ for the blue component. This underproduction of lanthanide-poor material could, however, be a consequence of neglecting neutrino absorption in the outflows. 
The blue component material has larger velocities (see histograms in Fig.~\ref{fig:v_r_out}) such that it punches through and quickly overtakes the nearly isotropic envelope of red component, which could otherwise obscure it (see Fig.~\ref{fig:kilonova_snapshots} and supplementary videos).

\section*{Acknowledgements} 
IMC thanks Dr.~K.~Alexander, Dr.~W.~Fong, and Dr.~B.~Metzger for their supportive discussions. 
RF acknowledges support from the Natural Sciences and Engineering Research Council of Canada (NSERC) through Discovery Grant RGPIN-2017-04286, and from the Faculty of Science at the University of Alberta. This work was supported in part by a Simons Investigator award from the Simons Foundation (EQ) and the Gordon and Betty Moore Foundation through Grant GBMF5076. This was was supported by NASA through grant 80NSSC18K0565 (FF, AT). This research used resources of the National Energy Research Scientific Computing Center (NERSC), which is supported by the Office of Science of the U.S. Department of Energy under Contract No. DE-AC02-05CH11231. The software used in this work was in part developed by the DOE NNSA-ASC OASCR Flash Center at the University of Chicago. Computations were performed at Carver, Hopper, and Edison (repositories m1186, m2058, m2401, and the scavenger queue).


\bibliographystyle{mnras} 
\bibliography{bib.bib} 

\appendix
\label{lastpage}
\end{document}